\documentclass[11pt,a4paper]{article}

\usepackage[utf8]{inputenc}
\usepackage[T1]{fontenc}
\usepackage{lmodern}
\usepackage[margin=1in]{geometry}
\usepackage{amsmath,amssymb,amsfonts,amsthm}
\usepackage{graphicx}
\usepackage{booktabs}
\usepackage{tabularx}
\usepackage{multirow}
\usepackage{array}
\usepackage{float}
\usepackage{caption}
\usepackage{setspace}
\usepackage[expansion=false]{microtype}
\usepackage{xcolor}
\usepackage[authoryear,round]{natbib}
\usepackage{authblk}
\usepackage{adjustbox}
\usepackage{xurl}
\usepackage{seqsplit}
\usepackage{lineno}
\usepackage[colorlinks=true,breaklinks=true,linkcolor=blue!60!black,citecolor=blue!60!black,urlcolor=blue!60!black]{hyperref}

\newcommand{\code}[1]{\nolinkurl{#1}}

\newtheorem{proposition}{Proposition}

\title{\textbf{Beyond Invoice Volumes: Temporal Network Structure, Maturity Concentration, and Systemic Liquidity Exposure in SME Trade-Credit Networks}}

\author[1]{Sai Srikanth Madugula\thanks{Corresponding Author. Email: saisrikanth.madugula\_phd.2023@woxsen.edu.in | ORCID: \href{https://orcid.org/0000-0001-6479-8443}{0000-0001-6479-8443}}}
\author[2]{Peplluis Esteva de la Rosa\thanks{Email: joseplluis.delarosa@udg.edu | ORCID: \href{https://orcid.org/0000-0003-1412-4170}{0000-0003-1412-4170}}}
\author[3]{Daya Shankar\thanks{Email: daya.shankar@woxsen.edu.in}}

\affil[1]{\small School of Technology, Woxsen University, Hyderabad, Telangana 502345, India}
\affil[2]{\small Universitat de Girona, 17004 Girona, Spain}
\affil[3]{\small School of Sciences, Woxsen University, Hyderabad, Telangana 502345, India}

\date{}

\begin{document}
\maketitle

\begin{abstract}
\noindent Trade-credit invoices encode directed contractual obligations whose temporal synchronization and counterparty concentration are routinely obscured by annual volume aggregates. This study conducts a comprehensive, leakage-free empirical and computational evaluation of 749,952 atomic B2B invoices issued between 2012 and 2023---representing EUR 99.705 billion in recorded face value across 75,343 analytical firms---drawn from the open pseudonymized Mendeley Data repository \citep{Esteva2026Mendeley}. Rigorous provenance validation separates the 2012--2023 analytical cohort from 1,240 terminal 2024 bridge records (EUR 630.12 million; 9 bridge-only firms) and isolates 13 extreme-duration contracts ($>3{,}650$ days; EUR 297.75 thousand) for explicit sensitivity auditing. Using directed weighted graphs, Herfindahl--Hirschman concentration indices (HHI), Shannon entropy, top-counterparty shares (CR1), spectral centralities, and multi-cadence (daily to annual) temporal operators, we formulate and benchmark a Maturity-Aware Network Exposure (MANE) indicator. Face values exhibit severe concentration: the top 1\% of invoices account for 81.41\% of total face value ($\text{Gini} = 0.9742$). Two-way fixed-effects panel models with circular date-score block bootstrapping ($B=999$, 12-month blocks) and Holm--Bonferroni multiplicity control reveal that higher lagged debtor counterparty concentration is significantly associated with lower maturity-aware exposure ($\hat{\beta} = -0.494$, 95\% CI $[-0.752, -0.248]$, $p_{\text{Holm}} = 0.004$) after scale and maturity controls. Four extended robustness batteries (R1--R4) confirm that this association survives 2021 separate-pipeline exclusion ($\hat{\beta} = -0.402$, $p = 0.004$), Shannon entropy ($\hat{\beta} = +0.257$, $p = 0.001$) and CR1 ($\hat{\beta} = -0.515$, $p = 0.001$) substitution, 1\%--99\% Winsorization ($\hat{\beta} = -0.462$), and restriction to multi-invoice debtors ($\text{count} \ge 2$, $\hat{\beta} = -0.334$, $p = 0.031$). Conversely, within-fixed-effects variance-inflation diagnostics reveal severe multicollinearity in joint-concentration models ($\text{VIF} = 22.22$), whose positive sensitivity vanishes among multi-invoice debtors ($p = 0.719$), while sub-period splitting uncovers a striking structural sign reversal in macro weighted-reciprocity dynamics between 2012--2019 ($\hat{\beta} = -0.706$, $p = 0.001$) and 2020--2023 ($\hat{\beta} = +0.232$, $p = 0.001$). In 12-month purged out-of-sample forecasting of an internal contractual-maturity target, adding MANE does not yield a statistically robust error reduction over direct maturity seeds alone ($\Delta\text{MSE} = 2.668$, 95\% CI $[-0.027, 7.954]$; trimmed $\Delta\text{MSE} = 0.067$, 95\% CI $[-0.035, 0.194]$). Our findings position MANE as a transparent, reproducible diagnostic for contractual screening and data-quality auditing---rather than a standalone proxy for realized default---and establish rigorous engineering requirements for supply-chain finance, multilateral clearing, and real-world-asset (RWA) tokenization pipelines.

\medskip
\noindent\textbf{Keywords:} Trade credit; Temporal networks; Maturity concentration; Systemic liquidity exposure; Counterparty risk; Complex financial networks; Supply chain finance; Explainable fintech.
\end{abstract}

\section{Introduction and Theoretical Motivation}
\label{sec:intro}

Interfirm trade credit represents one of the largest sources of short-term external financing for small and medium-sized enterprises (SMEs), embedding both a directed debtor--creditor claim and an explicit maturity schedule within every commercial invoice. Conventional portfolio reporting aggregates these invoices into annual face-value totals. Yet annual volume summaries cannot distinguish a diversified obligations book spread smoothly across counterparties and settlement dates from an identically sized portfolio clustered onto a single anchor counterparty or a narrow maturity window. Foundational corporate-finance literature links SME reliance on supplier credit to credit rationing in institutional lending markets \citep{Petersen1997}, while contract-theoretic models emphasize the lower diversion risk of physical goods relative to cash and show that bank credit and trade credit can operate as either complements or substitutes \citep{Burkart2004}. Accordingly, dense trade-credit linkages cannot be treated a priori as purely fragile or purely stabilizing.

Contractual maturities also reflect asymmetric bargaining and working-capital dynamics. Empirical studies associate weaker supplier bargaining power with expanded credit provision and extended payment terms \citep{Fabbri2016}, and connect receivables and inventory conversion cycles to firm profitability \citep{Deloof2003}. Although contractual invoice ledgers alone do not observe realized cash settlements, balance-sheet liquidity buffers, or formal bankruptcy events, they enable a rigorous computational inquiry: \emph{to what extent do directed topology, temporal maturity alignment, and counterparty--maturity concentration expose structural liquidity patterns that aggregate invoice volumes conceal?}

Recent advances in computational supply-chain finance---spanning multilateral invoice clearing \citep{Esteva2026Atomic,Esteva2026Multilateral}, ex-ante supplier payable-relief estimation \citep{Esteva2026ExAnte}, and explainable machine learning for under-collateralized decentralized SME credit pricing \citep{Madugula2026LeakageFree,Madugula2026DynamicLoan,Madugula2026RKAMM}---underscore the urgent need for auditable, leakage-free network diagnostics built directly on atomic invoice graphs. Supply-chain finance encompasses both bank-intermediated receivables financing and broader interfirm working-capital coordination \citep{Gelsomino2016}; evaluating the contractual invoice layer with strict separation between topological exposure and realized default is a prerequisite before operational deployment.

In this paper, we perform an independent, end-to-end empirical and computational audit of the publicly released Mendeley Data master invoice corpus \citep{Esteva2026Mendeley} (DOI: \href{https://doi.org/10.17632/28rbmvwsm9.1}{\nolinkurl{10.17632/28rbmvwsm9.1}}, CC BY 4.0) and its companion software architecture (\url{https://github.com/peplluis7/onemillioninvoices}). Our contributions are fourfold:
\begin{enumerate}
    \item \emph{Cryptographically Auditable Data Reconciliation:} We independently validate exact integer-cent totals, SHA-256 integrity, schema invariants, and cohort boundaries across all 751,192 raw records, isolating the 749,952 analytical 2012--2023 invoices (75,343 firms) from 1,240 terminal 2024 bridge records (9 bridge-only firms) and quantifying the exact impact of 13 extreme-duration contracts ($>3{,}650$ days).
    \item \emph{Maturity-Aware Network Exposure (MANE) Formulation:} Building on dynamic Katz resolvents and time-ordered walk operators, we formalize directional snapshot and recursive temporal MANE indicators with provable spectral convergence, scale equivariance, and upper bounds.
    \item \emph{Multiplicity-Controlled Econometrics and Extended R1--R4 Robustness Audits:} We evaluate four operational hypotheses (H1--H4) via two-way fixed-effects panels with circular date-score block bootstrapping ($B=999$) and Holm--Bonferroni family-wise error control, reinforced by four targeted diagnostic batteries: (R1) 2021 separate-pipeline cohort exclusion at both snapshot and raw-ledger levels, (R2) within-fixed-effects variance-inflation factor (VIF) collinearity diagnostics, (R3) 2012--2019 versus 2020--2023 structural regime splits uncovering a macro sign reversal in H3, and (R4) Shannon entropy, CR1 top-counterparty share, 1\%--99\% Winsorization, and multi-invoice debtor filtering ($\text{count} \ge 2$).
    \item \emph{Leakage-Free Out-of-Sample Benchmarking and Privacy-Safe Reporting:} Adopting strict 12-month temporal purging aligned with leakage-free financial ML standards \citep{Madugula2026LeakageFree}, we test whether MANE improves forward prediction of contractual maturity pressure beyond direct maturity seeds, while further masking entity identifiers (\code{Entity-01}--\code{Entity-10}) to mitigate graph-linkage re-identification risks.
\end{enumerate}

\section{Literature Review and Critical Comparison}
\label{sec:lit}

\subsection{Financial Interconnectedness Versus Realized Loss}
Network analyses of interfirm credit consistently demonstrate that structural position carries information beyond firm size. Studies of Korean interfirm credit networks relate PageRank centrality to modeled liquidity-shortage propagation \citep{Lee2016} and identify central buyers as key drivers of potential systemic risk \citep{Kwon2018}. Using Italian bank transaction records, \citet{Berloco2021} combine firm-level balance-sheet features with trade-credit network spillover metrics to substantially improve corporate bankruptcy recall. Crucially, however, modeled liquidity shortages, topological centrality, and realized corporate defaults represent distinct estimands that must not be conflated.

Theoretical financial-network models reinforce this caution. \citet{Acemoglu2015} establish a robust-yet-fragile phase transition in which dense interconnections absorb small shocks through diversification yet amplify systemic contagion once shock magnitudes exceed a critical threshold. \citet{Eboli2025} demonstrates that centralized interbank topologies contain localized distress but experience abrupt systemic breakdown when core hubs face simultaneous stress. Empirically, \citet{Asgharian2022} find in syndicated loan networks that network centrality primarily acts as a moderator amplifying firm-specific risk rather than an unconditional standalone risk factor, while agent-based supply-network simulations show that financial ripple effects depend jointly on network topology, exposure weights, and bargaining terms \citep{Proselkov2023}.

Clearing and multilateral settlement studies introduce explicit equilibrium and feasibility mechanisms beyond pure graph topology. \citet{Eisenberg2001} formalize simultaneous clearing payments under limited liability and debt priority; \citet{Hayakawa2019} links sequential settlement liquidity demands to directed graph structure; and \citet{Gaffeo2018} show that multilateral netting can reduce total defaults while redistributing residual losses across creditors. Recent work by our group directly models multilateral invoice clearing on temporal invoice graphs---comparing path-enabled compensation against bounded-cycle netting under common-day causal scheduling \citep{Esteva2026Multilateral,Esteva2026Atomic}---and estimates ex-ante payable relief for supplier selection \citep{Esteva2026ExAnte}. The present study complements rather than duplicates those clearing interventions: it audits the underlying contractual exposure geometry without claiming realized clearing savings or unobserved default outcomes.

\subsection{Directionality, Edge Weights, and Structural Centralities}
Classical network science distinguishes distinct structural dimensions of node centrality from global network centralization \citep{Freeman1978} and demonstrates that exchange power can diverge from simple degree centrality \citep{Bonacich1987}. In weighted socio-economic networks, link intensity and topological connectivity jointly govern structural heterogeneity \citep{Barrat2004}, motivating weighted centrality generalizations that balance tie count against transaction volume \citep{Opsahl2010}. Consequently, undirected gross-volume rankings conflate debtor liabilities with creditor receivables and obscure directional bottlenecks.

Path-based and spectral metrics likewise carry nuanced operational semantics. While \citet{Brandes2001} enables scalable shortest-path betweenness computation on sparse graphs, topological shortest paths in an invoice network do not automatically correspond to legally enforceable settlement routes. Similarly, PageRank---originally engineered for hypertext authority ranking \citep{Brin1998}---serves here strictly as a directed structural comparator. Surveys of complex network architecture emphasize joint evaluation of degree distributions, clustering, and assortativity \citep{Newman2003}; directed triangle motifs require orientation-aware clustering \citep{Fagiolo2007}, and weighted motif intensity can overturn unweighted topological conclusions \citep{Onnela2005}.

\subsection{Temporal Structure, Concentration, and Mesoscale Organization}
Static aggregation of time-stamped interactions frequently induces time-violating paths that are causally infeasible in the underlying event sequence \citep{Holme2012}. Time-ordered graphs generalize centrality to evolving edges \citep{Kim2012}, and dynamic communicability employs time-ordered matrix products to enforce causal directionality \citep{Grindrod2011}. Crucially, \citet{Colman2016} prove that temporal walk-based centralities blend temporal ordering with static topological density, requiring explicit decoupling against static baselines.

Attenuated multi-hop walk scoring traces back to \citet{Katz1953}, extended into aging-discounted dynamic matrix iterations \citep{Grindrod2013} and dynamic Katz resolvent formulations \citep{Arrigo2022}. Recent temporal-betweenness work further confirms that active versus passive walk definitions substantially alter node intermediacy rankings \citep{Naima2025}. Our MANE operator belongs squarely within this dynamic resolvent family rather than claiming a disjoint mathematical paradigm. Unlike two-layer B2B network evolution and monetary-flow models calibrated to Japanese transaction data \citep{Ozaki2024}, our dataset captures contractual issue--due windows rather than settled cash flows, making explicit sensitivity benchmarking indispensable.

At the portfolio and mesoscale levels, counterparty concentration (HHI) and maturity-bin concentration capture orthogonal dimensions of liquidity vulnerability, complemented by Shannon entropy as an allocation-diversity benchmark \citep{Shannon1948}. Weighted reciprocity similarly distinguishes balanced bidirectional flows from asymmetric binary reciprocal links \citep{Squartini2013}. For community detection, Louvain modularity optimization \citep{Blondel2008} offers high scalability on large sparse graphs, though it can occasionally produce poorly connected clusters---a limitation addressed by the Leiden refinement \citep{Traag2019}. Finally, $k$-core decomposition identifies nested minimum-degree shells \citep{Seidman1983}, which differs conceptually from formal core--periphery blockmodels \citep{Borgatti2000}; reporting both diagnostics prevents misinterpreting hub-dominated stars as cohesive economic cores.

\subsection{Supply-Chain Finance, Explainable AI Credit Scoring, and Tokenization}
Contemporary empirical finance highlights sharp directional asymmetries in trade credit. \citet{Gu2026} find that enterprise blockchain implementation reduces received trade credit from suppliers while expanding trade credit provided to customers under demand uncertainty. \citet{Yang2025} similarly uncover asymmetric industry peer contagion in trade-credit decisioning. In corporate debt studies, \citet{Zhang2024} document a nonlinear relationship between supplier concentration and debt maturity structure, and \citet{Li2024} show that supply-chain finance mitigates investment--financing maturity mismatches. Meanwhile, \citet{Grigoris2022} tie trade-credit counterparty risk and relationship duration to cross-sectional asset pricing, and a recent systematic review of multilayer financial networks warns that single-layer analyses can miss substantial cross-channel amplification when cash, bank credit, and portfolio overlaps are unobserved \citep{Scattergood2026}.

Crucially, digital infrastructure alone does not neutralize structural liquidity risk. Critical reviews of blockchain supply-chain finance emphasize persistent legal, governance, and data-oracle bottlenecks \citep{Ioannou2022}. Analytical models confirm that smart-contract adoption \citep{Wang2022} and deep-tier invoice tokenization \citep{Chen2026} yield liquidity- and regime-dependent value rather than unconditional gains. Practitioner analyses \citep{Sangari2025}, empirical comparisons of successful versus failed blockchain supply-chain finance initiatives \citep{Sriraman2026}, and multi-platform case studies \citep{Jia2026} consistently stress alignment between ledger architecture and invoice-level information-processing frictions.

This insight directly bridges our empirical invoice-network study with recent decentralized credit-underwriting architectures. \citet{Madugula2026LeakageFree} demonstrate that rigorous out-of-sample credit-risk pipelines must eliminate cross-validation data leakage, enforce probabilistic calibration, and explicitly audit privacy vulnerabilities such as membership inference on SME borrower records. Complementarily, \citet{Madugula2026DynamicLoan} and \citet{Madugula2026RKAMM} integrate SHAP-interpretable risk scores with a Reverse Kelly Automated Market Maker (rkAMM) for dynamic loan pricing on real-world assets. Grounded in this emerging paradigm, the present study supplies the missing empirical upstream layer: testing whether temporal invoice-graph concentration and maturity synchronization yield robust, leakage-free contractual screening signals before underwriting or tokenization.

\section{Research Questions and Operational Hypotheses}
\label{sec:hypotheses}

To rigorously evaluate structural liquidity exposure beyond raw invoice volumes, we operationalize four two-sided empirical hypotheses alongside descriptive topology, rank-stability, extended R1--R4 diagnostics, and purged out-of-sample predictive benchmarks:
\begin{itemize}
    \item \emph{H1 (Counterparty Concentration $\rightarrow$ Exposure):} Lagged debtor counterparty concentration ($\text{HHI}^{\text{counter,out}}_{i,t-1}$) is conditionally associated with current $\log(1 + \text{MANE}_{i,t})$, controlling for firm and month fixed effects, lagged gross scale, lagged debtor invoice count, and lagged value-weighted remaining maturity.
    \item \emph{H2 (Role Balance $\rightarrow$ Maturity Concentration):} Lagged debtor--creditor role balance ($B_{i,t-1}$) is conditionally associated with current debtor seven-day maturity-bin concentration ($\text{HHI}^{\text{mat,out}}_{i,t}$). We explicitly distinguish this firm-level role-balance proxy from a structural core-versus-periphery contrast, which we examine separately via retrospective $k$-core stratification.
    \item \emph{H3 (Macro Topology Dynamics $\rightarrow$ System Maturity Concentration):} Monthly first differences in system-wide weighted reciprocity ($\Delta r_{w,t}$) are associated with monthly first differences in system-wide value-weighted debtor maturity HHI ($\Delta \overline{\text{HHI}}^{\text{mat,out}}_t$), adjusting for active node/edge growth and annual seasonality.
    \item \emph{H4 (Joint Concentration $\rightarrow$ Exposure):} Lagged joint counterparty--maturity concentration ($\text{HHI}^{\text{joint,out}}_{i,t-1}$) is conditionally associated with current $\log(1 + \text{MANE}_{i,t})$ after controlling for both marginal HHIs ($\text{HHI}^{\text{counter,out}}_{i,t-1}$ and $\text{HHI}^{\text{mat,out}}_{i,t-1}$) and scale controls.
\end{itemize}
All four primary $p$-values constitute a single inference family adjusted via the sequentially rejective Holm--Bonferroni procedure \citep{Holm1979}, controlling family-wise Type-I error without assuming independence among tests.

\section{Data Provenance, Validation, and Sample Construction}
\label{sec:data}

All empirical quantities in this study are computed directly from the atomic master dataset \code{pseudonymized_atomic_invoice_records_2012_2024-02-08.csv} (independently verified decompressed SHA-256: \texttt{\seqsplit{e139b8dc75b7334d243874b6ebdaf1798c9969038b193f7eb4cd0e873bcc1a88}}), published in Mendeley Data V1 under a CC BY 4.0 license \citep{Esteva2026Mendeley} with companion curation and scheduling code archived at \url{https://github.com/peplluis7/onemillioninvoices}. The underlying records originate from pseudonymized commercial invoice filings sourced via the Ministry of Economics (IMI), Romania, and contain eight normalized fields: \code{invoice_code}, \code{debtor_code}, \code{creditor_code}, integer \code{amount_cents}, ISO \code{issue_date}, ISO \code{due_date}, integer \code{issue_cohort}, and \code{analysis_role} (\code{issue_cohort} vs.\ \code{terminal_bridge_only}). Annual year-split files from the repository were reconciled row-for-row against the master on all identifiers and substantive fields and were never concatenated to prevent double counting.

Validation enforced strict computational invariants prior to graph construction: string-typed firm and invoice identifiers, positive integer euro-cent amounts summed in arbitrary-precision Python integers (verified to fit within signed 64-bit integers prior to sparse matrix construction), zero missing values, zero duplicate invoice IDs, zero self-invoices ($d_i \neq c_i$), zero negative contractual terms ($u_i \ge s_i$), and exact alignment between \code{issue_date} year and \code{issue_cohort}. Calendar dates were mapped to exact integer epoch days relative to 1970-01-01 to avoid 64-bit nanosecond timestamp overflow caused by extreme recorded due dates (maximum recorded due date: 5656-03-08).

Two curation characteristics govern interpretation \citep{Esteva2026Mendeley,Esteva2026Atomic}. First, 58,239 analytical records belong to economic-duplicate groups sharing identical debtor, creditor, face value, issue date, and due date under distinct \code{invoice_code} identifiers; consistent with atomic accounting, all distinct invoice codes are retained. Second, the 2021 corpus was curated via a separate upstream extraction pipeline without a publicly released entity crosswalk, and the dataset has no pre-2012 opening ledger state; inter-year shifts therefore reflect both economic dynamics and administrative coverage variation, motivating our explicit R1 (2021 exclusion) and R3 (sub-period regime split) audits. Finally, to minimize graph-linkage disclosure risk beyond repository pseudonymization, all firm-level highlights and Top-10 rankings reported in this manuscript replace repository codes with further-masked analytical identifiers (\code{Entity-01}--\code{Entity-10}, \code{Entity-PR1}).

\section{Mathematical Formulation and Feature Engineering}
\label{sec:math}

\subsection{Directed Contractual Graphs and Topological Diagnostics}
Let each atomic invoice $i$ specify debtor $d_i$, creditor $c_i$, positive integer amount $a_i$ (euro cents), issue day $s_i$, and due day $u_i$, with contractual term $L_i = u_i - s_i \ge 0$. For an observation date $t$ and rolling lookback $\ell$ (with $\ell = 0$ denoting full observed history), the \emph{issued} invoice set is $\mathcal{I}^{\text{iss}}(t,\ell) = \{i : t - \ell + 1 \le s_i \le t\}$, and the \emph{scheduled} active set is $\mathcal{I}^{\text{sch}}(t,\ell) = \{i \in \mathcal{I}^{\text{iss}}(t,\ell) : u_i \ge t\}$. Scheduled membership represents an active contractual window rather than verified unpaid status.

The directed integer-cent adjacency matrix is
\begin{equation}
W_{vw}(t,\ell) = \sum_{i \in \mathcal{I}(t,\ell)} a_i \,\mathbf{1}(d_i = v,\; c_i = w).
\end{equation}
For each active firm $v$, outgoing (debtor) strength is $s_v^{\text{out}} = \sum_w W_{vw}$, incoming (creditor) strength is $s_v^{\text{in}} = \sum_w W_{wv}$, gross incident value is $G_v = s_v^{\text{out}} + s_v^{\text{in}}$ (noting that summing $G_v$ across all nodes counts each invoice twice), and role balance is
\begin{equation}
B_v = \frac{2 \min\!\left(s_v^{\text{out}}, s_v^{\text{in}}\right)}{G_v} \in [0, 1].
\end{equation}
Directed density is $m/[n(n-1)]$, and weighted reciprocity follows \citet{Squartini2013}:
\begin{equation}
r_w = \frac{\sum_{v \neq w} \min(W_{vw}, W_{wv})}{\sum_{v \neq w} W_{vw}}.
\end{equation}
Adjacent snapshot edge stability is evaluated via weighted Jaccard persistence:
\begin{equation}
J_w(t) = \frac{\sum_{v,w} \min\!\left(W^{(t)}_{vw},\, W^{(t-1)}_{vw}\right)}{\sum_{v,w} \max\!\left(W^{(t)}_{vw},\, W^{(t-1)}_{vw}\right)}.
\end{equation}

Conventional structural comparators include directed incoming PageRank on $W$ and outgoing PageRank on $W^\top$ (damping $0.85$, uniform teleportation, dangling redistribution over active nodes), shortest-path weighted betweenness with edge distance $100/W_{vw}$ (exact for $n \le 5{,}000$; 64-source sampled for $n > 5{,}000$), largest strongly connected component (SCC) eigenvector centrality (validated via scaled ARPACK solver with relative residual $< 10^{-14}$), directed weighted clustering \citep{Fagiolo2007}, directed weighted Louvain modularity \citep{Blondel2008}, undirected $k$-core shell numbers \citep{Seidman1983}, and a restricted degree-ordered binary core--periphery template maximizing Pearson correlation with an ideal hub-incident template \citep{Borgatti2000}.

\subsection{Counterparty, Maturity-Bin, and Joint Concentration}
For firm $v$ with positive directional volume in role $r \in \{\text{out}, \text{in}\}$, let $p_{v,w}^{r}$ denote the face-value share allocated to counterparty $w$. Counterparty concentration is measured via three complementary metrics: the Herfindahl--Hirschman index $\text{HHI}_{v}^{\text{counter},r} = \sum_w (p_{v,w}^r)^2$, the top-counterparty concentration ratio $\text{CR1}_{v}^{\text{counter},r} = \max_w p_{v,w}^r$, and Shannon allocation entropy $H_v^{\text{counter},r} = -\sum_w p_{v,w}^r \log p_{v,w}^r$ \citep{Shannon1948}. Partitioning remaining contractual maturity into discrete bins $b_i = \lfloor (u_i - t)/h \rfloor$ (primary width $h = 7$ days; sensitivities $h \in \{1, 30\}$ days) defines maturity-bin concentration $\text{HHI}_v^{\text{mat},r} = \sum_b (p_{v,b}^r)^2$ and joint counterparty--maturity concentration $\text{HHI}_v^{\text{joint},r} = \sum_{w,b} (p_{v,w,b}^r)^2$. Single-counterparty or single-bin active portfolios mechanically attain $\text{HHI} = 1$, $\text{CR1} = 1$, and $H = 0$; inactive roles are encoded as missing rather than zero.

\subsection{Maturity-Aware Network Exposure (MANE)}
To encode contractual proximity to maturity, we define euro-scaled directional maturity seeds with exponential urgency decay parameter $\tau > 0$:
\begin{equation}
q_v^{\text{out}}(t) = \sum_{i : d_i = v} \frac{a_i}{100} \exp\!\left(-\frac{u_i - t}{\tau}\right), \qquad
q_v^{\text{in}}(t) = \sum_{i : c_i = v} \frac{a_i}{100} \exp\!\left(-\frac{u_i - t}{\tau}\right).
\end{equation}
Let $C_t^{\text{out}}$ row-normalize $W(t,\ell)$ and $C_t^{\text{in}}$ row-normalize $W(t,\ell)^\top$, preserving zero rows so each operator is row-substochastic ($\|C_t^r\|_\infty \le 1$). The static snapshot resolvent comparator is
\begin{equation}
M_t^r = (1 - \alpha)\left(I - \alpha C_t^r\right)^{-1} q_t^r, \qquad r \in \{\text{out}, \text{in}\},\quad \alpha \in [0, 1).
\end{equation}
Across ordered observation dates $t_1 < t_2 < \dots < t_J$, temporal MANE updates recursively with memory decay $\gamma > 0$:
\begin{equation}
z_{t_j}^r = (1 - a_j)\, q_{t_j}^r + a_j \, C_{t_j}^r z_{t_{j-1}}^r, \qquad a_j = \alpha \exp\!\left(-\frac{t_j - t_{j-1}}{\gamma}\right), \qquad z_{t_1}^r = q_{t_1}^r.
\end{equation}
Combined temporal MANE is $T_{v,t} = (z_{v,t}^{\text{out}} + z_{v,t}^{\text{in}})/2$, with primary calibration $\alpha = 0.5$, $\tau = 30$ days, and $\gamma = 90$ days.

\begin{proposition}[Spectral Existence, Boundedness, and Scale Equivariance]
For any row-substochastic matrix $C_t^r$ and $\alpha \in [0,1)$: (i) $(I - \alpha C_t^r)^{-1} = \sum_{k=0}^{\infty} \alpha^k (C_t^r)^k$ converges absolutely; (ii) both $M_t^r$ and $z_{t_j}^r$ are elementwise nonnegative and satisfy $\|M_t^r\|_\infty \le \|q_t^r\|_\infty$ and $\|z_{t_j}^r\|_\infty \le \max_{1 \le m \le j} \|q_{t_m}^r\|_\infty$; and (iii) scaling all invoice face values by $c > 0$ scales $M_t^r$, $z_{t_j}^r$, and $T_{v,t}$ identically by $c$ while leaving $C_t^r$ invariant.
\end{proposition}
\noindent\emph{Proof sketch.} See Appendix~\ref{app:math}. Because $C_t^r$ averages neighbor states, an individual node's $T_{v,t}$ can exceed its own gross face value $G_v$ when linked to high-seed neighbors; thus MANE captures contractual network-contextual exposure rather than conserved monetary loss.

\section{Statistical Methodology and Extended Robustness Design}
\label{sec:methods}

Monthly scheduled panels are constructed across rolling 30-, 90-, and 365-day windows plus full observed history, complemented by daily 90-day scalar topology/MANE tracks and quarterly/annual matching-date comparisons. For firm-level hypotheses (H1, H2, H4), we estimate two-way fixed-effects models restricted to strictly consecutive observed calendar-month lags:
\begin{equation}
y_{i,t} = \mu_i + \lambda_t + \beta \, x_{i,t-1} + \boldsymbol{\theta}^\top \mathbf{X}_{i,t-1} + \varepsilon_{i,t},
\end{equation}
where $\mu_i$ and $\lambda_t$ denote firm and month fixed effects absorbed via alternating projections, and $\mathbf{X}_{i,t-1}$ controls for lagged $\log(1 + \text{gross EUR})$, lagged $\log(1 + \text{debtor invoice count})$, and lagged value-weighted remaining maturity (with H4 additionally conditioning on lagged marginal counterparty and maturity HHIs). For macro hypothesis H3, monthly first differences in value-weighted debtor maturity HHI are regressed on first-differenced weighted reciprocity, $\Delta \log(1 + n_t)$, $\Delta \log(1 + m_t)$, and annual sine/cosine seasonal harmonics ($N = 143$ months).

To account for serial correlation and arbitrary within-date cross-firm correlation without distorting fixed-effect projections, inference employs circular whole-date score-block bootstrapping ($B = 999$ draws, primary block length $L = 12$ months; sensitivities $L \in \{6, 18\}$ months) alongside date-aggregated HAC standard errors \citep{Kunsch1989,Politis1994,Cameron2011}. Beyond the baseline block-length, maturity-bin, and extreme-term sensitivities, we execute four independent structural diagnostics directly on the raw master CSV:
\begin{itemize}
    \item \emph{R1 (2021 Separate-Pipeline Exclusion):} Re-estimating H1--H4 both after excluding 2021 snapshot dates from the panel (R1a) and after dropping all \code{issue_cohort == 2021} invoices from the raw ledger and rebuilding the entire 2012--2023 temporal network from scratch (R1b).
    \item \emph{R2 (Within-FE Multicollinearity Audit):} Computing exact variance inflation factors (VIF) and condition numbers on the within-demeaned design matrices of H1, H2, H4, and H4+seeds to diagnose collinearity between joint and marginal HHIs.
    \item \emph{R3 (Sub-Period Structural Regime Split):} Estimating H1--H4 separately across the early low-volume regime (2012--2019; 95 dates) and the high-volume surge regime (2020--2023; 48 dates).
    \item \emph{R4 (Alternative Concentration, Winsorization, and Multi-Invoice Filtering):} Testing H1 with Shannon counterparty entropy (R4a) and CR1 top-counterparty share (R4b), applying 1\%--99\% Winsorization to continuous variables (R4c), and restricting firm panels to genuinely multi-invoice debtors ($\text{count}^{\text{out}}_{i,t-1} \ge 2$, R4d) to rule out mechanical single-invoice $\text{HHI} = 1$ boundary effects.
\end{itemize}

For out-of-sample predictive benchmarking, we predict next-month $\log(1 + q_{i,t+1}^{\text{out}})$ using expanding training windows separated from each held-out calendar year (2016--2023; 95 held-out dates) by a strict 12-month temporal purge, fitting all scaling and preprocessing transformations exclusively on training partitions to eliminate look-ahead leakage \citep{Madugula2026LeakageFree}. Nested models successively add volume/maturity controls, conventional centrality, concentration metrics, current maturity seeds, and temporal MANE, benchmarked against a zero-parameter persistence baseline. Equal-date MSE differences between the maturity-seed and seed-plus-MANE models are evaluated via 12-month block intervals.

\section{Empirical Results}
\label{sec:results}

\subsection{Sample Reconciliation and Heavy-Tailed Distribution}
Table~\ref{tab:audit} summarizes the independently audited provenance and distribution profile of the master corpus \citep{Esteva2026Mendeley}. Separating the 1,240 terminal 2024 bridge records removes 9 bridge-only firms from the 75,352 full-stream firm union, establishing an exact analytical denominator of 75,343 firms and 749,952 invoices totaling EUR 99.705 billion (9,970,501,556,950 cents). Invoice face values display extreme right skewness: the median invoice is EUR 1,272.35 versus a mean of EUR 132,948.53 (SD EUR 5.39 million; max EUR 1.921 billion), yielding a Gini coefficient of 0.9742 and a top-1\% face-value share of 81.41\%. In accordance with \citet{Clauset2009}, we characterize this heavy tail via exact quantiles and concentration shares rather than asserting an unvalidated parametric power law. Median contractual duration is 10 days (99th percentile: 268 days), with 71,123 same-day contracts (9.48\%) and 13 extreme-duration contracts exceeding 3,650 days (total EUR 297,751.10; 0.0003\% of analytical value).

\begin{table}[htbp]
\centering
\caption{Master Dataset Provenance Audit and Distributional Summary (2012--2024). Computed directly from \citet{Esteva2026Mendeley}.}
\label{tab:audit}
\small
\adjustbox{max width=\textwidth}{\begin{tabular}{@{}lrr@{}}
\toprule
\textbf{Audit / Distributional Metric} & \textbf{Analytical (2012--2023)} & \textbf{Bridge (2024)} \\
\midrule
Invoice records ($N$) & 749,952 & 1,240 \\
Total face value (exact EUR cents) & 9,970,501,556,950 & 63,011,692,269 \\
Total face value (EUR billions) & 99.705 & 0.630 \\
Unique firms (bridge-only in col.\ 2) & 75,343 & 9 (75,352 total) \\
Integrity violations (missing / dup ID / self / neg.\ term) & 0 & 0 \\
Economic-duplicate records (distinct IDs retained) & 58,239 & --- \\
Same-day contractual terms ($L_i = 0$ days) & 71,123 (9.48\%) & --- \\
Extreme terms ($L_i > 3{,}650$ days; EUR face value) & 13 (EUR 297,751.10) & --- \\
Face value: Median / Mean / SD (EUR) & 1,272.35 / 132,948.53 / 5,389,195.79 & --- \\
Face value: 95th / 99th / Max (EUR) & 147,080.08 / 1,239,060.41 / 1.921B & --- \\
Face value inequality: Gini / Top-1\% value share & 0.9742 / 81.41\% & --- \\
Contractual term: Q25 / Median / Q75 / Q99 (days) & 10.0 / 10.0 / 30.0 / 268.0 & --- \\
\bottomrule
\end{tabular}}
\end{table}

\begin{figure}[htbp]
\centering
\includegraphics[width=0.86\textwidth]{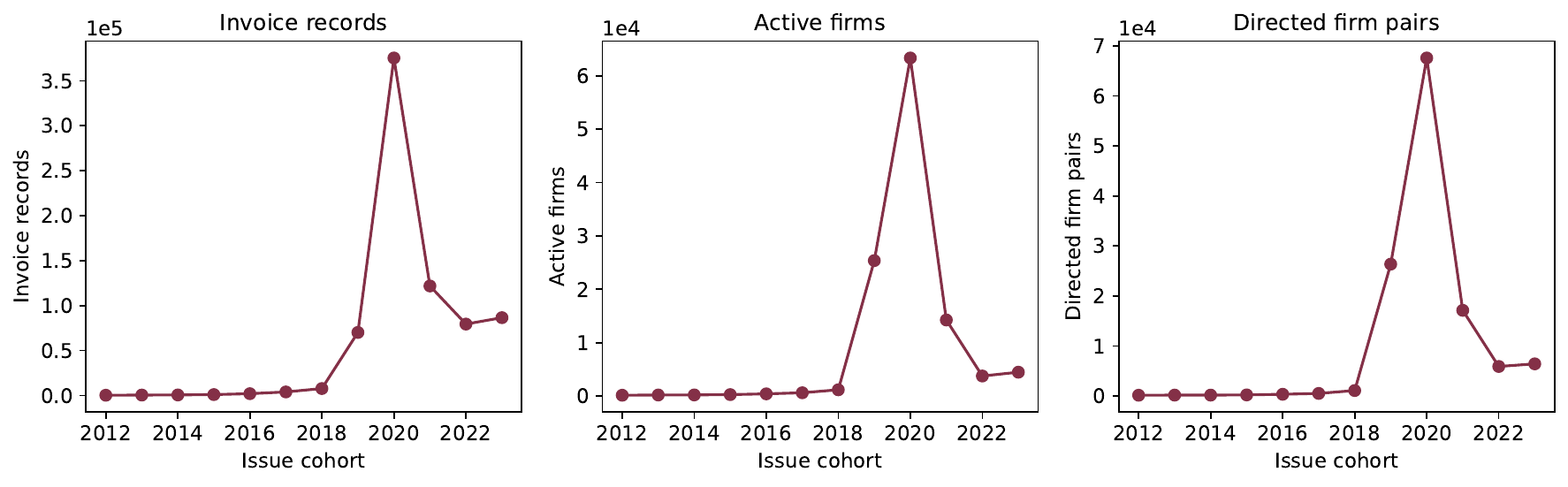}
\caption{Annual analytical issue-cohort invoice counts, active firms, and directed pairs (2012--2023; 2024 bridge excluded). Source: \citet{Esteva2026Mendeley}.}
\label{fig:annual}
\end{figure}

Annual issued-cohort topology (Table~\ref{tab:annual} and Fig.~\ref{fig:annual}) demonstrates pronounced non-stationarity in administrative coverage and value density: 2020 accounts for 50.05\% of all analytical invoices (375,386 records across 63,371 firms), whereas 2022 captures 41.70\% of total face value (EUR 41.576 billion across just 3,743 firms) alongside peak annual weighted reciprocity (0.633).

\begin{table}[htbp]
\centering
\caption{Annual Issued-Cohort Trade-Credit Network Structure (2012--2023). Reciprocity is value-weighted; 2024 bridge records excluded.}
\label{tab:annual}
\small
\adjustbox{max width=\textwidth}{\begin{tabular}{@{}crrrrr@{}}
\toprule
\textbf{Year} & \textbf{Invoices} & \textbf{Active Firms} & \textbf{Directed Pairs} & \textbf{Value (EUR bn)} & \textbf{Weighted Reciprocity} \\
\midrule
2012 & 339 & 138 & 129 & 0.069 & 0.008 \\
2013 & 540 & 184 & 158 & 0.119 & 0.000 \\
2014 & 666 & 196 & 153 & 0.197 & 0.006 \\
2015 & 1,069 & 256 & 196 & 0.538 & 0.004 \\
2016 & 2,078 & 400 & 323 & 0.679 & 0.015 \\
2017 & 4,004 & 601 & 491 & 0.257 & 0.040 \\
2018 & 7,798 & 1,160 & 1,070 & 0.460 & 0.077 \\
2019 & 70,155 & 25,383 & 26,358 & 4.424 & 0.151 \\
2020 & 375,386 & 63,371 & 67,600 & 14.020 & 0.470 \\
2021 & 121,844 & 14,250 & 17,116 & 15.153 & 0.413 \\
2022 & 79,471 & 3,743 & 5,883 & 41.576 & 0.633 \\
2023 & 86,602 & 4,460 & 6,409 & 22.212 & 0.493 \\
\bottomrule
\end{tabular}}
\end{table}

\subsection{Role Asymmetry, Concentration, and Mesoscale Architecture}
\begin{figure}[htbp]
\centering
\includegraphics[width=0.78\textwidth]{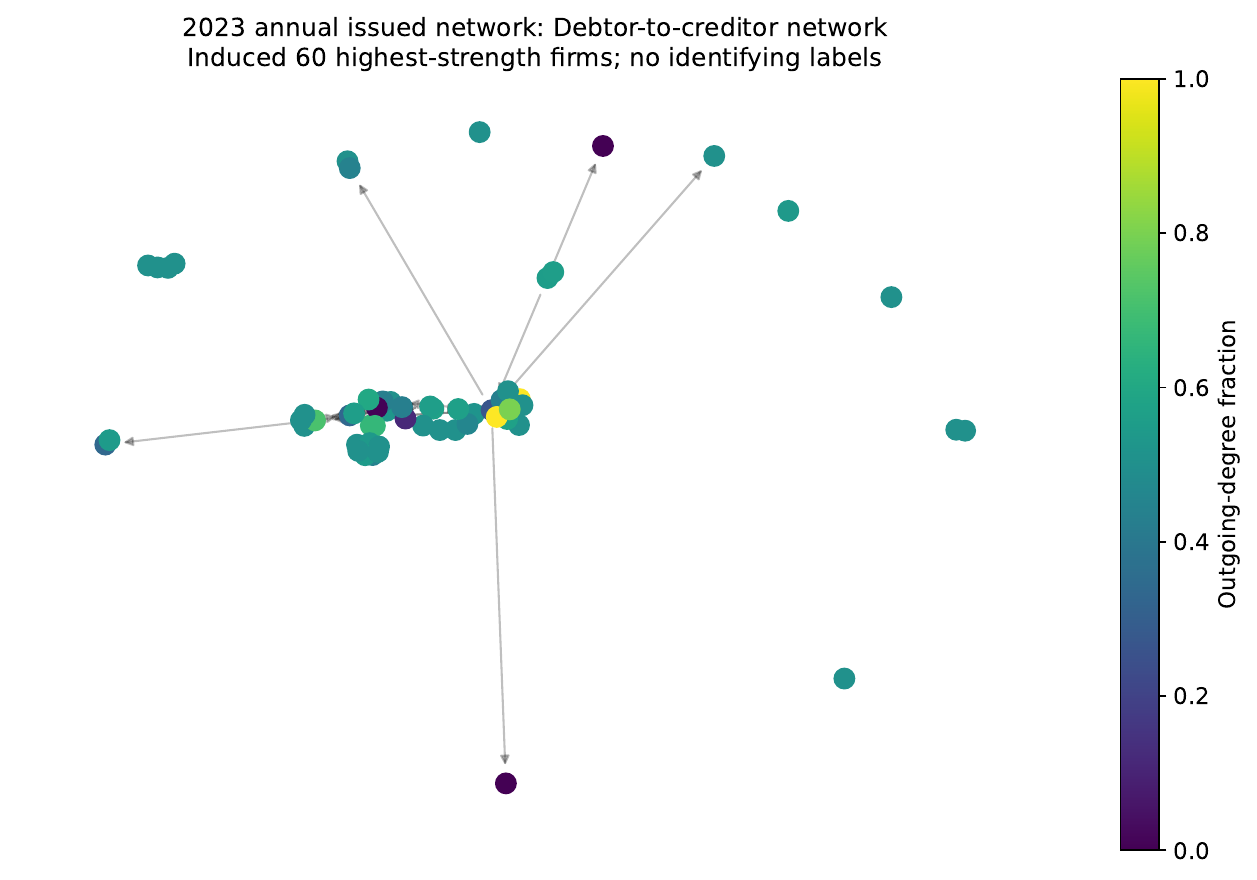}
\caption{Directed 2023 issued-cohort subgraph induced by the top 60 firms by total incident face value. Edges point debtor $\to$ creditor; node color encodes outgoing degree share.}
\label{fig:net}
\end{figure}

\begin{figure}[htbp]
\centering
\includegraphics[width=0.86\textwidth]{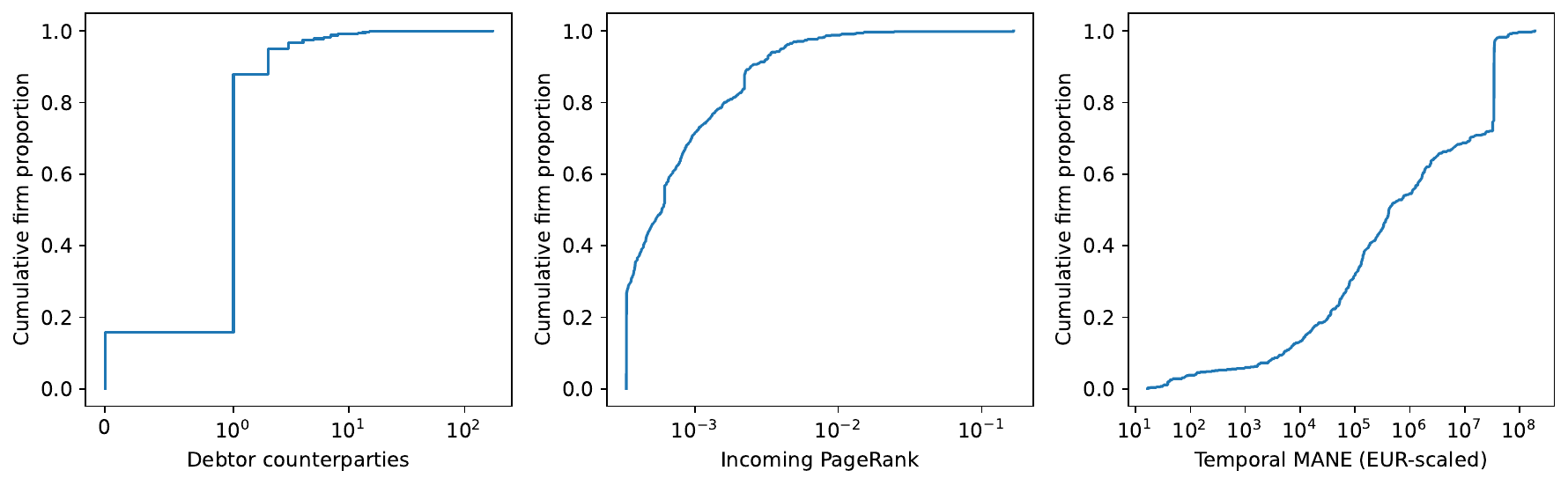}
\caption{Distributions of debtor counterparty count, incoming PageRank, and temporal MANE across active firms in the 31 December 2023 90-day scheduled snapshot.}
\label{fig:cent}
\end{figure}

\begin{figure}[htbp]
\centering
\includegraphics[width=0.84\textwidth]{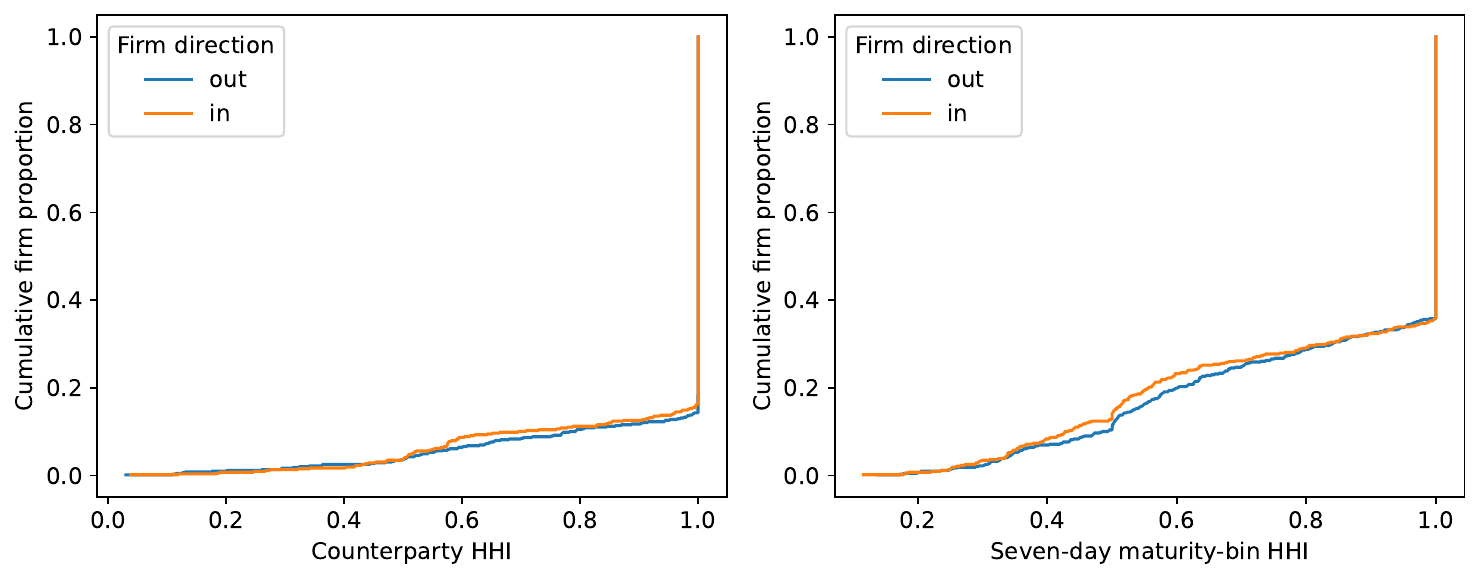}
\caption{Firm-level counterparty and seven-day maturity-bin HHI distributions by debtor (outgoing) and creditor (incoming) roles on 31 December 2023.}
\label{fig:conc}
\end{figure}

On the terminal 31 December 2023 90-day scheduled snapshot, 4,526 invoices connect 698 active firms across 971 directed pairs, with 405 firms holding dual debtor--creditor roles. Among the 587 active debtors, 85.52\% exhibit counterparty $\text{HHI} = 1.0$ (mean $0.947$, SD $0.160$) and 64.22\% exhibit seven-day maturity-bin $\text{HHI} = 1.0$ (mean $0.855$, SD $0.231$); comparable distributions appear among the 516 active creditors (counterparty HHI mean $0.942$; maturity HHI mean $0.845$). Crucially, these high firm-level HHI shares largely reflect thin active-snapshot portfolios (median outgoing invoice count $= 1.0$, 75th percentile $= 3.75$) rather than pervasive insolvency risk (Figs.~\ref{fig:net}--\ref{fig:conc}).

\begin{figure}[htbp]
\centering
\includegraphics[width=0.78\textwidth]{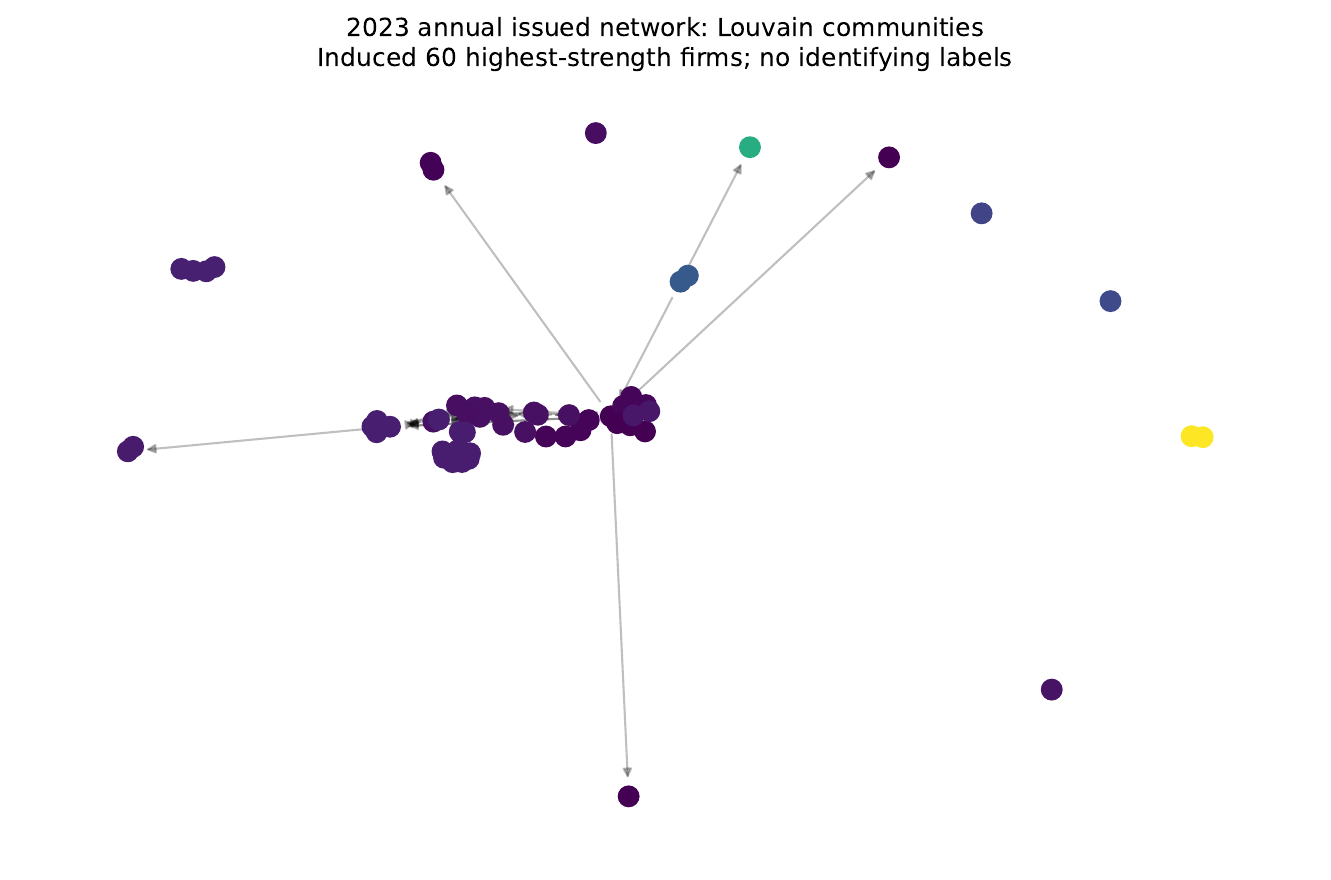}
\caption{Louvain community labels on the induced 60-firm 2023 subgraph. Algorithmic partitions reflect transactional clustering rather than legal ownership.}
\label{fig:comm}
\end{figure}

\begin{figure}[htbp]
\centering
\includegraphics[width=0.78\textwidth]{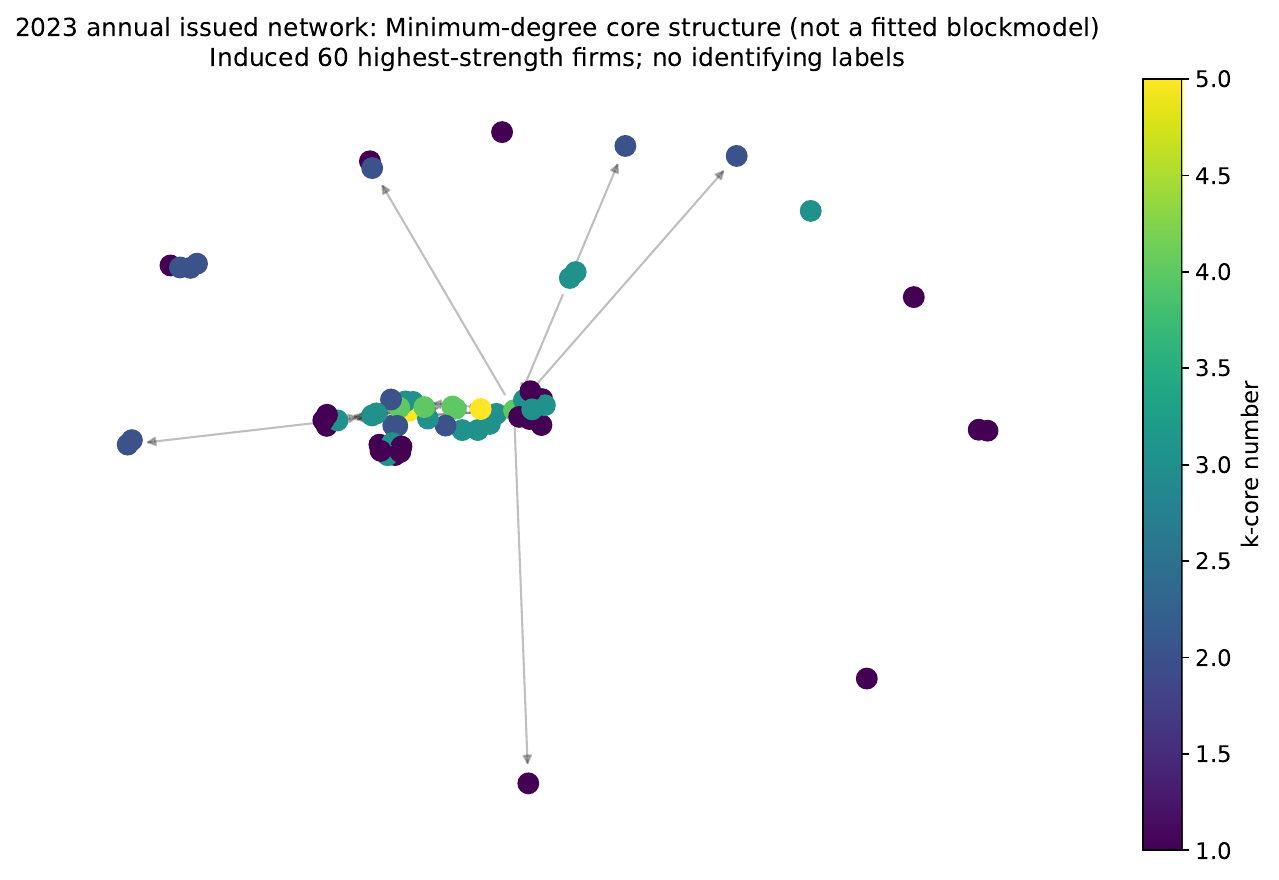}
\caption{Undirected $k$-core shell numbers on the induced 60-firm 2023 subgraph.}
\label{fig:kcore}
\end{figure}

Across the full 2023 annual issued cohort ($n = 4{,}460$), Louvain partitioning yields 300 communities with modularity $Q = 0.703$, an undirected maximum $k$-core of 5 comprising 16 firms, a largest weakly connected component (WCC) spanning 81.10\% of firms, and a largest strongly connected component (SCC) containing 27.85\% ($1{,}242$ firms; Figs.~\ref{fig:comm}--\ref{fig:kcore}). Strikingly, fitting the degree-ordered binary core--periphery template selects a 3-firm core whose \emph{core--core density is exactly zero}---illustrating that hub-centric star motifs can achieve positive template correlation without forming an internally cohesive economic core.

ARPACK spectral decomposition of the 2023 largest SCC (relative residual $7.79 \times 10^{-16}$, strictly nonnegative) places maximum eigenvector mass on \code{Entity-01} and \code{Entity-02}, which also occupy Ranks 1 and 2 on 31 December 2023 temporal MANE, whereas incoming PageRank peaks at \code{Entity-PR1} ($0.1671$). Table~\ref{tab:top10} presents the further-masked Top-10 temporal MANE entities, revealing how network propagation elevates low-volume nodes attached to systemically concentrated counterparties: \code{Entity-10}, with just EUR 0.16M in gross incident value, registers EUR 61.62M in temporal MANE---a $374.80\times$ score-to-gross ratio arising entirely from neighbor-seeded exposure.

\begin{table}[htbp]
\centering
\caption{Further-Masked Top-10 Firms Ranked by Temporal MANE (31 December 2023, 90-Day Scheduled Snapshot). Monetary figures in EUR millions; raw repository pseudonyms are masked to mitigate graph-linkage disclosure risk.}
\label{tab:top10}
\small
\adjustbox{max width=\textwidth}{\begin{tabular}{@{}clrrrrrcc@{}}
\toprule
\textbf{Rank} & \textbf{Masked ID} & \textbf{Gross (M)} & \textbf{MANE (M)} & \textbf{$q^{\text{out}}$ (M)} & \textbf{$q^{\text{in}}$ (M)} & \textbf{Role Bal.} & \textbf{$d^{\text{out}}$} & \textbf{$d^{\text{in}}$} \\
\midrule
1 & \code{Entity-01} & 706.27 & 189.41 & 78.88 & 490.70 & 0.252 & 10 & 5 \\
2 & \code{Entity-02} & 596.64 & 166.59 & 471.28 & 0.00 & 0.000 & 1 & 0 \\
3 & \code{Entity-03} & 105.81 & 94.57 & 83.88 & 10.39 & 0.382 & 7 & 3 \\
4 & \code{Entity-04} & 46.13 & 90.58 & 1.95 & 43.12 & 0.130 & 1 & 1 \\
5 & \code{Entity-05} & 2.25 & 73.21 & 1.87 & 0.17 & 0.150 & 2 & 1 \\
6 & \code{Entity-06} & 13.88 & 66.08 & 0.00 & 13.88 & 0.000 & 0 & 1 \\
7 & \code{Entity-07} & 10.50 & 64.99 & 0.00 & 10.50 & 0.000 & 0 & 1 \\
8 & \code{Entity-08} & 0.43 & 61.71 & 0.00 & 0.36 & 0.000 & 0 & 1 \\
9 & \code{Entity-09} & 0.32 & 61.64 & 0.00 & 0.14 & 0.000 & 0 & 1 \\
10 & \code{Entity-10} & 0.16 & 61.62 & 0.00 & 0.06 & 0.000 & 0 & 1 \\
\bottomrule
\end{tabular}}
\end{table}

\subsection{Temporal Evolution and Ranking Divergences}
\begin{figure}[htbp]
\centering
\includegraphics[width=0.86\textwidth]{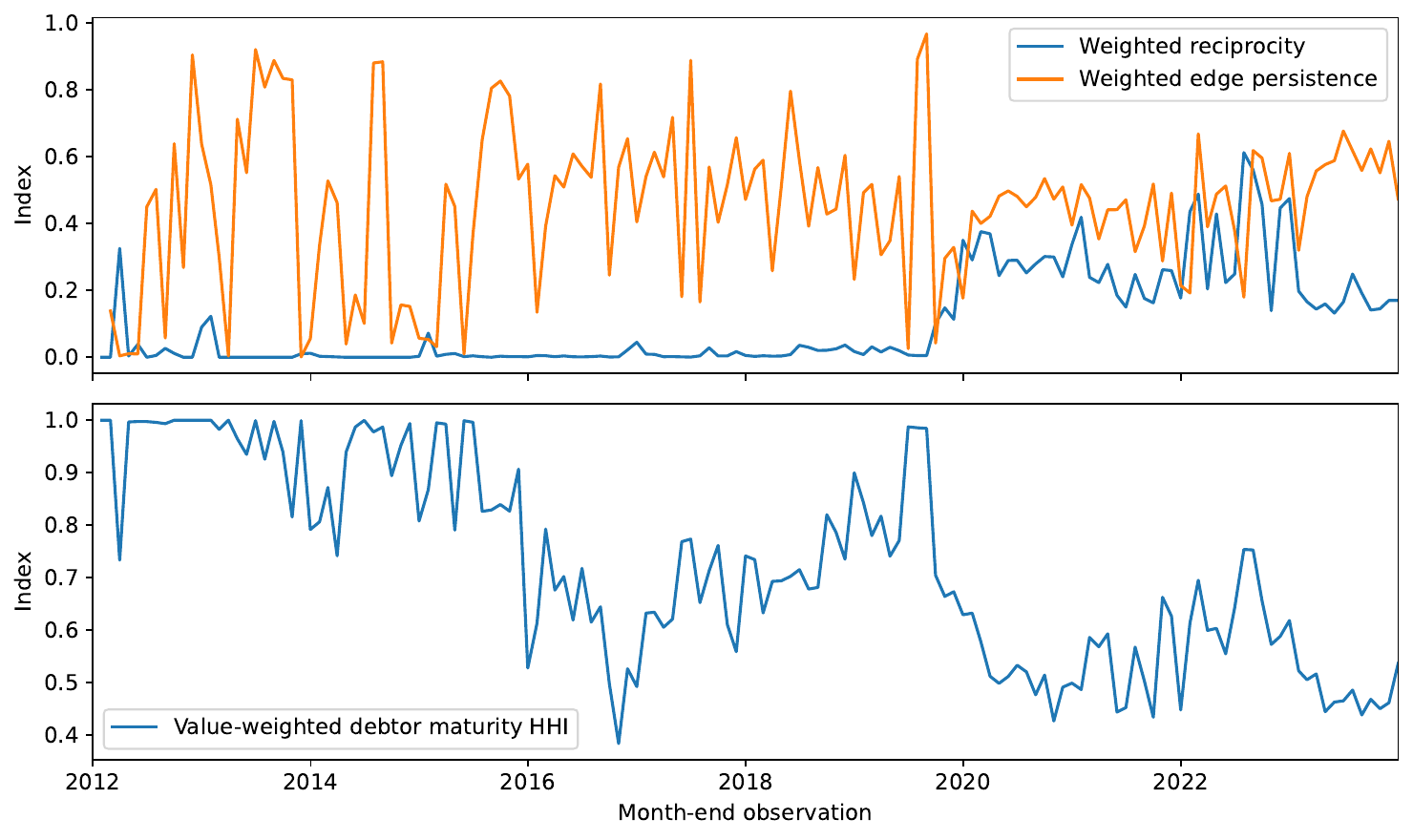}
\caption{Monthly 90-day scheduled-snapshot weighted reciprocity, weighted edge persistence, and value-weighted debtor maturity HHI (2012--2023).}
\label{fig:temp}
\end{figure}

\begin{figure}[htbp]
\centering
\includegraphics[width=0.66\textwidth]{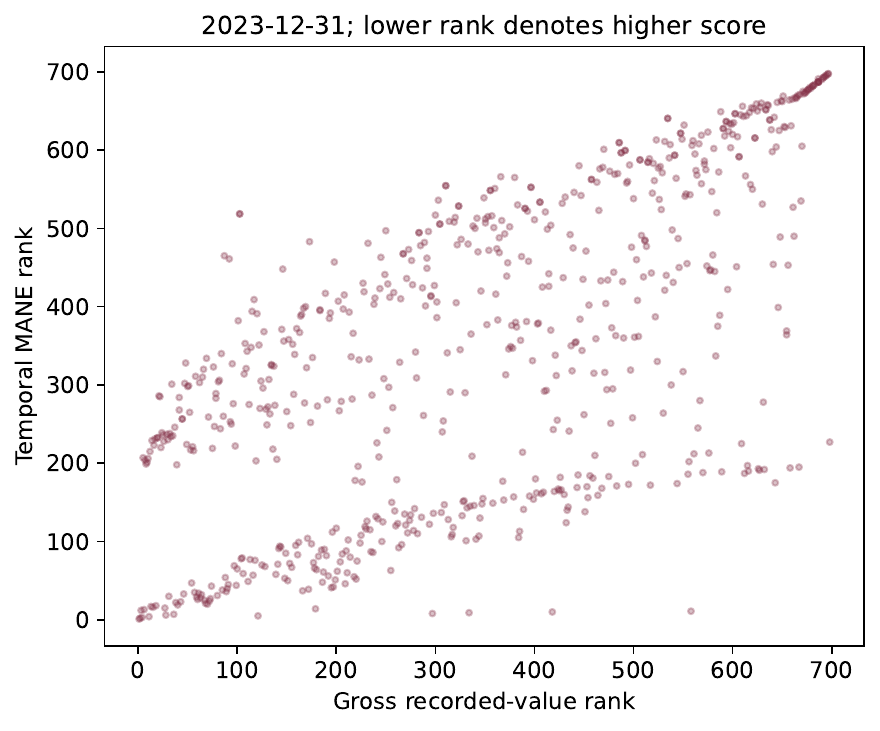}
\caption{Comparison of firm ranks by gross incident value versus temporal MANE on 31 December 2023 ($n=698$; lower rank denotes higher exposure).}
\label{fig:rank}
\end{figure}

In the terminal 90-day snapshot, system-level weighted reciprocity stands at $0.170$, month-over-month weighted edge Jaccard persistence at $0.473$ (partly mechanical due to overlapping 90-day windows), and value-weighted debtor maturity HHI at $0.536$ (Fig.~\ref{fig:temp}). Across the 698 active firms (Fig.~\ref{fig:rank}), temporal MANE correlates moderately with gross incident value ($\rho = 0.665$), weakly with incoming PageRank ($\rho = 0.343$), role balance ($\rho = 0.343$), and direct outgoing seed ($\rho = 0.286$), negatively with debtor maturity HHI ($\rho = -0.163$, $n=587$), and near-orthogonally with debtor counterparty HHI ($\rho = 0.040$, $n=587$). Conversely, temporal MANE tracks the static snapshot resolvent closely ($\rho = 0.969$), indicating that recursive temporal carry-over refines rather than completely overturns single-date resolvent rankings at the terminal horizon.

\subsection{Primary Hypothesis Tests and Extended R1--R4 Robustness Audits}
\begin{figure}[htbp]
\centering
\includegraphics[width=0.95\textwidth]{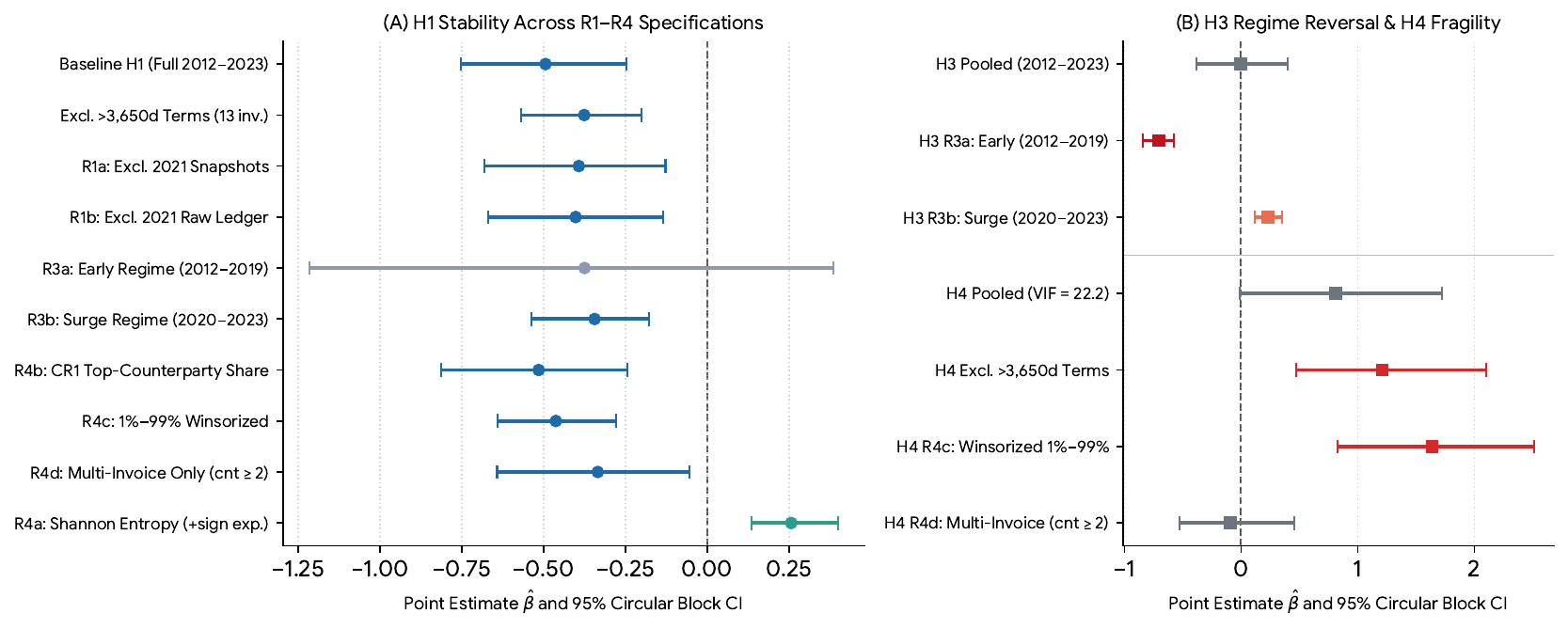}
\caption{Extended R1--R4 robustness summary. (A) H1 remains stably negative across 2021-cohort exclusion, regime splits, CR1 share, Winsorization, and multi-invoice filtering, flipping cleanly positive under Shannon diversity entropy. (B) Macro H3 undergoes a sharp sign reversal between 2012--2019 ($-0.706$) and 2020--2023 ($+0.232$), while collinear H4 ($\text{VIF}=22.2$) vanishes completely once single-invoice debtors are excluded.}
\label{fig:r1r4}
\end{figure}

Table~\ref{tab:hyp} reports the primary exploratory fixed-effects estimates, while Tables~\ref{tab:vif} and~\ref{tab:r1r4} (visualized in Fig.~\ref{fig:r1r4}) present the full baseline sensitivity and independent R1--R4 robustness batteries computed directly from the master CSV. Under H1, lagged debtor counterparty concentration exhibits a statistically significant negative association with $\log(1 + \text{MANE})$ ($\hat{\beta} = -0.494$, 95\% circular block CI $[-0.752, -0.248]$, $p_{\text{Holm}} = 0.004$, $N = 92{,}495$ firm-months across 15,523 firms and 143 dates): a $+0.10$ rise in lagged counterparty HHI corresponds to a $-0.0494$ shift in $\log(1 + \text{MANE})$, holding gross scale, invoice count, remaining maturity, and firm/date fixed effects constant.

\begin{table}[htbp]
\centering
\caption{Primary Hypothesis Tests (H1--H4) Using Two-Way Fixed Effects (H1, H2, H4) and Macro First Differences (H3). Inference uses 999 circular whole-date score-block bootstrap draws ($L=12$ months) and Holm--Bonferroni family adjustment.}
\label{tab:hyp}
\small
\adjustbox{max width=\textwidth}{\begin{tabular}{@{}lrrrrrr@{}}
\toprule
\textbf{Hypothesis \& Predictor} & \textbf{Estimate ($\hat{\beta}$)} & \textbf{95\% Block CI} & \textbf{HAC SE} & \textbf{Approx.\ $p$} & \textbf{$p_{\text{Holm}}$} & \textbf{Obs.} \\
\midrule
H1: Lagged debtor counterparty HHI & $-0.494$ & $[-0.752,\; -0.248]$ & 0.130 & 0.001 & \textbf{0.004} & 92,495 \\
H2: Lagged role balance & $0.0067$ & $[-0.0050,\; 0.0182]$ & 0.0058 & 0.236 & 0.472 & 89,032 \\
H3: $\Delta$ Weighted reciprocity & $-0.0049$ & $[-0.3850,\; 0.4027]$ & 0.2138 & 0.987 & 0.987 & 143 \\
H4: Lagged joint HHI & $0.812$ & $[-0.008,\; 1.721]$ & 0.471 & 0.081 & 0.243 & 92,495 \\
\bottomrule
\end{tabular}}
\end{table}

Neither H2 ($p_{\text{Holm}} = 0.472$) nor H3 ($p_{\text{Holm}} = 0.987$) receives support in the pooled full-sample specification. In a retrospective descriptive stratification on 31 December 2023, active debtors in the $k\text{-core} \ge 3$ shell ($n=66$, median MANE EUR 1.52M, dual-role share 73.7\%) showed lower mean outgoing maturity HHI ($0.770$) than peripheral $k\text{-core} = 1$ debtors ($n=418$, mean maturity HHI $0.877$, dual-role share 54.8\%); however, this descriptive contrast is unadjusted for confounding and leaves formal core--periphery causal inference open.

Our four extended diagnostic batteries (R1--R4) uncover three high-impact structural mechanisms that clarify every primary result:
\begin{enumerate}
    \item \emph{H1 Survives Every Structural Stress Test (R1, R3, R4):} Excluding the separately curated 2021 corpus leaves H1 robustly negative both when dropping 2021 snapshots (R1a: $\hat{\beta} = -0.392$, $p = 0.005$, $N = 76{,}222$) and when removing \code{issue_cohort == 2021} invoices at the raw ledger level and rebuilding the temporal network from scratch (R1b: $\hat{\beta} = -0.402$, $p = 0.004$, $N = 75{,}894$). Substituting alternative concentration measures confirms exact theoretical sign coherence: Shannon counterparty entropy (R4a, where higher values mean greater diversity) yields $\hat{\beta} = +0.257$ (95\% CI $[0.136, 0.400]$, $p = 0.001$), while CR1 top-counterparty share (R4b) reproduces HHI closely ($\hat{\beta} = -0.515$, 95\% CI $[-0.814, -0.244]$, $p = 0.001$). Applying 1\%--99\% Winsorization (R4c: $\hat{\beta} = -0.462$, $p = 0.001$) or restricting the panel strictly to multi-invoice debtors ($\text{count}^{\text{out}}_{i,t-1} \ge 2$, R4d: $\hat{\beta} = -0.334$, 95\% CI $[-0.642, -0.053]$, $p = 0.031$, $N = 39{,}106$) confirms that H1 is not a mechanical artifact of single-invoice $\text{HHI} = 1$ portfolios.
    \item \emph{Severe Within-FE Multicollinearity Explains H4 Fragility (R2 \& R4d):} Table~\ref{tab:vif} reveals pristine conditioning in H1 (max VIF $= 1.32$, condition number $= 1.72$) and H2 (max VIF $= 1.19$, condition number $= 1.56$), but severe within-FE multicollinearity in H4: lagged joint HHI attains $\text{VIF} = 22.22$ and lagged marginal maturity HHI attains $\text{VIF} = 19.65$ (condition number $= 10.92$; $11.64$ with seed controls). Because joint counterparty--maturity cells collapse onto maturity bins whenever a firm holds one or few counterparties, tiny tail perturbations swing H4 dramatically (rising to $+1.209$ under 13-record trimming and $+1.639$ under 1\%--99\% Winsorization). Crucially, once mechanical single-invoice debtors are excluded in R4d ($\text{count} \ge 2$), H4 collapses completely to null ($\hat{\beta} = -0.094$, 95\% CI $[-0.527, 0.460]$, $p = 0.719$), proving that positive H4 spikes are a collinearity and thin-portfolio artifact.
    \item \emph{Structural Regime Sign Reversal in Macro H3 (R3):} Splitting the timeline into the early sparse regime (2012--2019; 95 months) versus the post-2020 high-volume surge regime (2020--2023; 48 months) exposes a striking \emph{macro sign reversal} masked by the pooled H3 null ($-0.0049$, $p = 0.987$). During 2012--2019, increases in weighted reciprocity strongly \emph{reduced} system maturity concentration ($\hat{\beta} = -0.706$, 95\% CI $[-0.841, -0.577]$, $p = 0.001$), whereas during 2020--2023, increases in weighted reciprocity significantly \emph{raised} maturity concentration ($\hat{\beta} = +0.232$, 95\% CI $[+0.117, +0.354]$, $p = 0.001$). Meanwhile, H1 remains consistently negative across both eras ($-0.375$ in 2012--2019; $-0.344$, $p = 0.001$ in 2020--2023).
\end{enumerate}

\begin{table}[htbp]
\centering
\caption{R2 Within-Fixed-Effects Multicollinearity Diagnostics (Variance Inflation Factors and Design-Matrix Condition Numbers on Absorbed Firm--Month Panels).}
\label{tab:vif}
\small
\adjustbox{max width=\textwidth}{\begin{tabular}{@{}lrrrr@{}}
\toprule
\textbf{Within-FE Regressor} & \textbf{H1 Model} & \textbf{H2 Model} & \textbf{H4 Model} & \textbf{H4 + Seeds} \\
\midrule
Lagged debtor counterparty HHI & 1.157 & --- & 1.586 & 1.591 \\
Lagged role balance & --- & 1.047 & --- & --- \\
Lagged debtor maturity HHI & --- & --- & \textbf{19.648} & \textbf{19.650} \\
Lagged joint counterparty--maturity HHI & --- & --- & \textbf{22.218} & \textbf{22.243} \\
Lagged $\log(1 + \text{gross EUR})$ & 1.170 & 1.191 & 1.172 & 2.012 \\
Lagged $\log(1 + \text{debtor count})$ & 1.315 & 1.189 & 1.740 & 1.765 \\
Lagged value-weighted remaining maturity & 1.000 & 1.000 & 1.000 & 1.005 \\
Lagged $\log(1 + q^{\text{out}})$ / $\log(1 + q^{\text{in}})$ & --- & --- & --- & 1.761 / 1.263 \\
\midrule
\textbf{Condition Number ($\kappa$)} & 1.72 & 1.56 & \textbf{10.92} & \textbf{11.64} \\
\bottomrule
\end{tabular}}
\end{table}

\begin{table}[htbp]
\centering
\caption{Comprehensive Sensitivity and Extended R1--R4 Robustness Battery Across Baseline Perturbations, 2021-Cohort Exclusion (R1), Regime Splits (R3), and Concentration/Winsorization/Multi-Invoice Checks (R4). All CIs use 999 circular date-score block draws.}
\label{tab:r1r4}
\footnotesize
\adjustbox{max width=\textwidth}{\begin{tabular}{@{}llrrrrr@{}}
\toprule
\textbf{Hyp.} & \textbf{Specification / Diagnostic Test} & \textbf{$\hat{\beta}$} & \textbf{95\% Block CI} & \textbf{HAC SE} & \textbf{Approx.\ $p$} & \textbf{Obs.} \\
\midrule
\multicolumn{7}{@{}l}{\textit{Panel A: Baseline Block, Extreme-Term, and Bin-Width Sensitivities}} \\
H1 & Circular block $L = 6$ / $L = 18$ months & $-0.494$ & $[-0.771,-0.233]$ / $[-0.749,-0.215]$ & 0.140 / 0.139 & 0.004 / 0.001 & 92,495 \\
H1 & Exclude terms $> 3{,}650$ days (13 inv.) & $-0.376$ & $[-0.569,\; -0.200]$ & 0.092 & 0.001 & 92,488 \\
H4 & Exclude terms $> 3{,}650$ days (13 inv.) & $1.209$ & $[0.475,\; 2.101]$ & 0.448 & 0.008 & 92,488 \\
H4 & Add lagged direct maturity seeds & $1.248$ & $[0.453,\; 2.105]$ & 0.443 & 0.006 & 92,495 \\
\midrule
\multicolumn{7}{@{}l}{\textit{Panel B: R1 --- 2021 Separate-Pipeline Exclusion}} \\
H1 & R1a: Exclude 2021 snapshot dates & $-0.392$ & $[-0.681,\; -0.127]$ & 0.141 & 0.005 & 76,222 \\
H1 & R1b: Exclude 2021 raw cohort \& rebuild & $-0.402$ & $[-0.669,\; -0.135]$ & 0.134 & 0.004 & 75,894 \\
H2 & R1b: Exclude 2021 raw cohort \& rebuild & $0.0073$ & $[-0.0015,\; 0.0162]$ & 0.0044 & 0.105 & 73,199 \\
H3 & R1b: Exclude 2021 raw cohort \& rebuild & $-0.125$ & $[-0.554,\; 0.353]$ & 0.241 & 0.584 & 132 \\
H4 & R1b: Exclude 2021 raw cohort \& rebuild & $0.740$ & $[-0.209,\; 1.866]$ & 0.525 & 0.162 & 75,894 \\
\midrule
\multicolumn{7}{@{}l}{\textit{Panel C: R3 --- Structural Regime Split (2012--2019 vs.\ 2020--2023)}} \\
H1 & R3a: Early regime (2012--2019; 95 mos.) & $-0.375$ & $[-1.214,\; 0.387]$ & 0.417 & 0.367 & 12,725 \\
H1 & R3b: Surge regime (2020--2023; 48 mos.) & $-0.344$ & $[-0.536,\; -0.177]$ & 0.084 & 0.001 & 79,770 \\
H3 & R3a: Early regime (2012--2019; 95 mos.) & $\mathbf{-0.706}$ & $[-0.841,\; -0.577]$ & 0.104 & \textbf{0.001} & 95 \\
H3 & R3b: Surge regime (2020--2023; 48 mos.) & $\mathbf{+0.232}$ & $[0.117,\; 0.354]$ & 0.063 & \textbf{0.001} & 48 \\
H4 & R3b: Surge regime (2020--2023; 48 mos.) & $0.801$ & $[0.130,\; 1.805]$ & 0.424 & 0.059 & 79,770 \\
\midrule
\multicolumn{7}{@{}l}{\textit{Panel D: R4 --- Alternative Concentration, Winsorization, and Multi-Invoice Filter}} \\
H1 & R4a: Shannon counterparty entropy & $\mathbf{+0.257}$ & $[0.136,\; 0.400]$ & 0.067 & \textbf{0.001} & 92,495 \\
H1 & R4b: CR1 top-counterparty share & $\mathbf{-0.515}$ & $[-0.814,\; -0.244]$ & 0.148 & \textbf{0.001} & 92,495 \\
H1 & R4c: 1\%--99\% Winsorized continuous vars & $-0.462$ & $[-0.642,\; -0.278]$ & 0.089 & 0.001 & 92,495 \\
H2 & R4c: 1\%--99\% Winsorized continuous vars & $0.0078$ & $[-0.0047,\; 0.0195]$ & 0.0060 & 0.187 & 89,032 \\
H4 & R4c: 1\%--99\% Winsorized continuous vars & $1.639$ & $[0.830,\; 2.512]$ & 0.448 & 0.001 & 92,495 \\
H1 & R4d: Multi-invoice debtors ($\text{count} \ge 2$) & $\mathbf{-0.334}$ & $[-0.642,\; -0.053]$ & 0.155 & \textbf{0.031} & 39,106 \\
H2 & R4d: Multi-invoice debtors ($\text{count} \ge 2$) & $0.0052$ & $[-0.0059,\; 0.0141]$ & 0.0049 & 0.279 & 37,577 \\
H4 & R4d: Multi-invoice debtors ($\text{count} \ge 2$) & $-0.094$ & $[-0.527,\; 0.460]$ & 0.265 & 0.719 & 39,106 \\
\bottomrule
\end{tabular}}
\end{table}

\subsection{Parameter, Lookback, Cadence, and Purged Forward Prediction Audits}
\begin{figure}[htbp]
\centering
\includegraphics[width=0.82\textwidth]{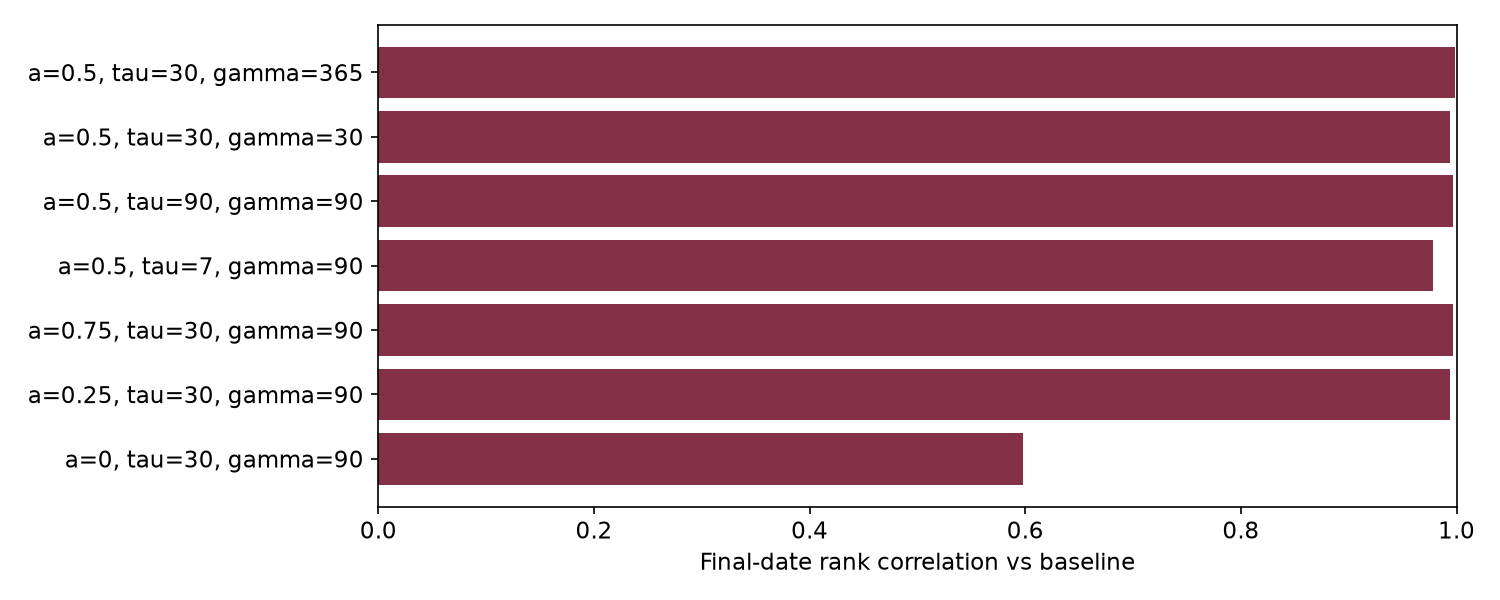}
\caption{Final-date Spearman rank correlations against baseline temporal MANE ($\alpha=0.5, \tau=30, \gamma=90$) across parameter perturbations ($n=698$).}
\label{fig:params}
\end{figure}

\begin{figure}[htbp]
\centering
\includegraphics[width=0.84\textwidth]{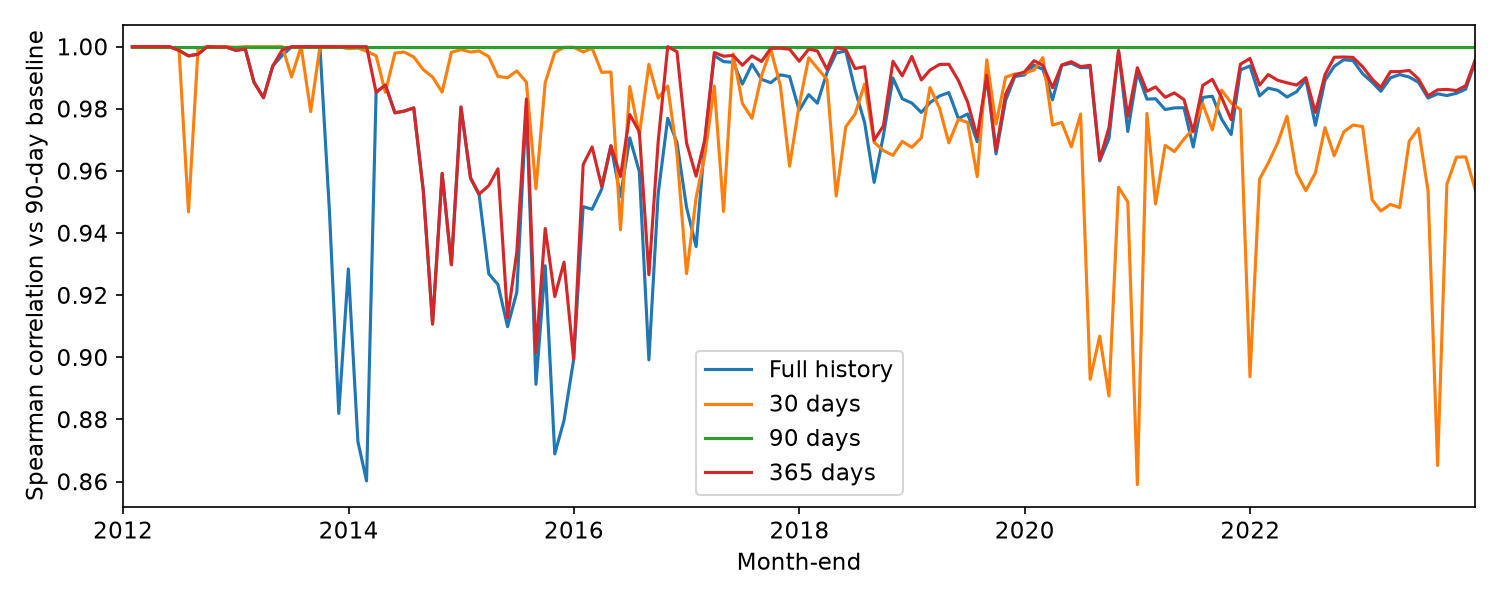}
\caption{Monthly common-firm Spearman rank stability across rolling 30-, 90-, and 365-day and full-history scheduled lookbacks.}
\label{fig:win}
\end{figure}

Across the parameter grid on 31 December 2023 ($n=698$, Fig.~\ref{fig:params}), temporal MANE rankings remain highly stable across non-zero damping and decay variations ($\alpha \in \{0.25, 0.75\} \Rightarrow \rho \in [0.994, 0.997]$; $\tau \in \{7, 90\} \Rightarrow \rho \in [0.979, 0.996]$; $\gamma \in \{30, 365\} \Rightarrow \rho \in [0.994, 0.999]$), whereas removing network propagation entirely ($\alpha = 0$) drops rank correlation sharply to $\rho = 0.598$. Across lookbacks (Fig.~\ref{fig:win}), the 365-day and full-history windows match the 90-day baseline closely ($\rho = 0.996$ and $0.995$), while the 30-day window yields $\rho = 0.954$ globally ($n=597$) yet only 76\% top-100 overlap---demonstrating that high global Spearman correlation can conceal material churn among top-ranked firms. Sampling cadence exerts an even stronger effect: daily-versus-monthly rank correlation drops to $-0.010$ at the 31 December 2021 boundary, whereas quarterly and annual cadences correlate with monthly MANE at $0.969$ and $0.899$ on 31 December 2023.

\begin{figure}[htbp]
\centering
\includegraphics[width=0.84\textwidth]{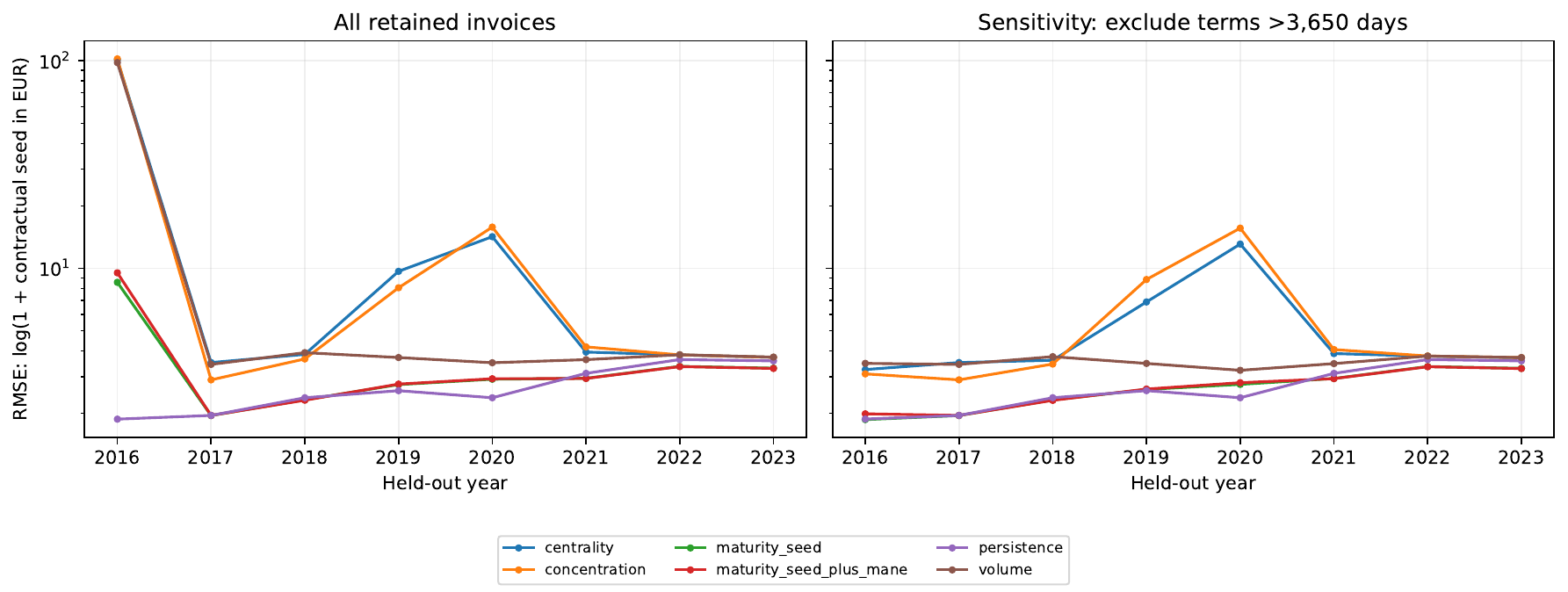}
\caption{Held-out-year prediction RMSE for next-month $\log(1+q^{\text{out}})$ with and without excluding the 13 extreme-duration contracts ($>3{,}650$ days), using a 12-month purge between training and test years.}
\label{fig:pred}
\end{figure}

Finally, Table~\ref{tab:pred} and Fig.~\ref{fig:pred} summarize the 12-month purged out-of-sample forecast evaluation of next-month $\log(1 + q_{i,t+1}^{\text{out}})$ across 95 held-out dates. Adding temporal MANE on top of direct maturity seeds yields an equal-date MSE difference of $+2.668$ (95\% block CI $[-0.027, 7.954]$), failing to outperform maturity seeds alone. Extreme untrimmed errors in the 2016 fold (volume RMSE $= 98.05$ vs.\ trimmed RMSE $= 3.48$) collapse upon removing the 13 extreme-duration records, shrinking the MSE difference to $+0.067$ (95\% CI $[-0.035, 0.194]$). Remarkably, simple zero-parameter persistence achieves competitive accuracy (2016 trimmed RMSE $= 1.878$; 2017 RMSE $= 1.955$), confirming that complex recursive graph features should be deployed strictly as structural screening diagnostics rather than presumed superior short-horizon forecasters.

\begin{table}[htbp]
\centering
\caption{Purged Out-of-Sample Forward Prediction Summary (Next-Month $\log(1+q^{\text{out}})$, 12-Month Temporal Purge, 95 Held-Out Dates). Positive $\Delta\text{MSE}$ indicates higher error for the augmented MANE model relative to direct maturity seeds alone.}
\label{tab:pred}
\small
\adjustbox{max width=\textwidth}{\begin{tabular}{@{}lrrrr@{}}
\toprule
\textbf{Evaluation Regime / Benchmark Comparison} & \textbf{Point Estimate} & \textbf{95\% Block CI} & \textbf{2016 Fold RMSE} & \textbf{2017 Fold RMSE} \\
\midrule
$\Delta\text{MSE}$ (Seed+MANE minus Seed, Untrimmed) & $+2.668$ & $[-0.027,\; 7.954]$ & 9.513 vs.\ 8.561 & 1.955 vs.\ 1.952 \\
$\Delta\text{MSE}$ (Seed+MANE minus Seed, Excl.\ $>3{,}650$d) & $+0.067$ & $[-0.035,\; 0.194]$ & 1.990 vs.\ 1.866 & 1.955 vs.\ 1.952 \\
Volume Baseline (Untrimmed / Trimmed) & --- & --- & 98.049 / 3.482 & 3.444 / 3.443 \\
Persistence Benchmark (Untrimmed / Trimmed) & --- & --- & 1.876 / 1.878 & 1.955 / 1.955 \\
\bottomrule
\end{tabular}}
\end{table}

\section{Discussion, Theoretical Contribution, and Practical Implications}
\label{sec:discussion}

Our findings yield four substantive insights for computational network science, SME finance, and decentralized supply-chain infrastructure. First, \emph{data provenance, population hygiene, and within-FE collinearity audits are first-order inferential safeguards.} Failing to separate the 2024 terminal bridge inflates the analytical firm population by nine firms; retaining just 13 anomalous due-date records ($0.0017\%$ of rows) inflates 2016 out-of-sample volume RMSE by $28\times$; and ignoring within-FE collinearity ($\text{VIF} = 22.22$) would misinterpret H4's positive spike under Winsorization as a true joint-concentration effect even though it completely vanishes among multi-invoice debtors (R4d: $p = 0.719$).

Second, \emph{directionality, maturity alignment, and neighbor propagation generate genuine structural separation from gross invoice volume ($\rho = 0.665$) with robust micro-level consistency.} The negative conditional association under H1 ($\hat{\beta} = -0.494$, confirmed across 2021-cohort exclusion, Shannon entropy $+0.257$, CR1 share $-0.515$, Winsorization $-0.462$, and multi-invoice filtering $-0.334$) demonstrates that, holding firm scale and fixed effects constant, diversified debtors embedded across multiple counterparties accumulate significantly greater neighbor-propagated maturity exposure than single-counterparty specialists. Conversely, zero core--core density within an apparently optimal binary core--periphery template cautions against equating hub-star heuristics with cohesive systemic cores \citep{Borgatti2000,Eboli2025}.

Third, \emph{macro-level network relationships are non-stationary across liquidity regimes.} Uncovering the opposite-signed H3 regime reversal---strongly negative in 2012--2019 ($\hat{\beta} = -0.706$) yet significantly positive in 2020--2023 ($\hat{\beta} = +0.232$)---shows how pooling heterogeneous coverage and macro-liquidity eras can induce Simpson's paradox at the network level. In sparse pre-2020 trading networks, reciprocal bilateral pairs corresponded to smoother staggered billing schedules, whereas in the dense, high-value 2020--2023 network (where annual reciprocity reached $0.633$), reciprocal commercial hubs synchronized contractual due dates onto common settlement windows.

Fourth, \emph{for SME supply-chain finance, multilateral clearing, and real-world-asset (RWA) tokenization}, contractual invoice graphs provide an auditable upstream screening layer that complements---but cannot replace---verified payment outcomes, cash reserves, and credit bureaus \citep{Scattergood2026}. In multilateral invoice clearing \citep{Esteva2026Atomic,Esteva2026Multilateral}, directed temporal graphs nominate candidate common-day clearing cycles and paths, yet actual liquidity relief depends on verifiable residual claims, legal enforceability, and participant consent. Likewise, in explainable AI-driven decentralized lending and dynamic loan pricing \citep{Madugula2026LeakageFree,Madugula2026DynamicLoan,Madugula2026RKAMM}, invoice-graph concentration and maturity-synchronization features must pass strict leakage-free validation, collinearity screening, anomaly-date filtering, and privacy-preserving disclosure audits before feeding on-chain pricing or collateralization rules.

\section{Limitations and Future Research}
\label{sec:limitations}

Several boundaries qualify these results. First, the Mendeley Data corpus \citep{Esteva2026Mendeley} reflects administrative filings with heterogeneous annual coverage, no pre-2012 opening stock, and a separately curated 2021 cohort lacking a public entity crosswalk; while R1 confirms H1's robustness to 2021 removal, temporal aggregates still combine economic evolution with administrative coverage changes. Second, contractual due dates do not record actual settlement timestamps, partial payments, disputes, credit notes, or corporate defaults, nor does the schema disclose firm size, industry sector, or banking facilities. Third, while circular date-score block bootstrapping handles time-series correlation and within-date cross-sectional dependence conditional on the observed panel, it does not replicate full graph-reconstruction uncertainty; additionally, large-graph annual betweenness relies on 64-source sampling, and Louvain clustering should be benchmarked against Leiden \citep{Traag2019} and stochastic core--periphery blockmodels in future work.

Promising avenues for subsequent research include: (i) linking contractual MANE diagnostics to external realized payment-delay and corporate bankruptcy registries; (ii) evaluating privacy-preserving graph embeddings under membership-inference threat models \citep{Madugula2026LeakageFree}; (iii) formalizing regime-switching models for the 2012--2019 versus 2020--2023 reciprocity--maturity transition discovered in R3; and (iv) integrating maturity-concentration signals into ex-ante multilateral clearing and RWA dynamic loan pricing frameworks \citep{Esteva2026ExAnte,Madugula2026RKAMM}.

\section{Conclusion}
\label{sec:conclusion}

Analyzing 749,952 atomic commercial invoices totaling EUR 99.705 billion across 75,343 firms demonstrates that trade-credit obligations exhibit extreme value skewness ($\text{Gini} = 0.9742$; top 1\% holding 81.41\% of value) and rich directed temporal structure obscured by annual volume totals. Across rigorous two-way fixed-effects panels with circular block inference and Holm--Bonferroni correction, higher lagged debtor counterparty concentration is robustly associated with lower maturity-aware network exposure ($\hat{\beta} = -0.494$, $p_{\text{Holm}} = 0.004$)---a relationship that survives 2021-cohort exclusion, Shannon entropy and CR1 substitution, 1\%--99\% Winsorization, and multi-invoice filtering. Conversely, our extended R2--R4 diagnostics prove that H4's positive sensitivity stems from severe within-FE multicollinearity ($\text{VIF} = 22.22$) and single-invoice boundary effects, while pooled macro H3 conceals a sharp regime sign reversal between 2012--2019 ($-0.706$) and 2020--2023 ($+0.232$). Because purged 12-month out-of-sample forecasting shows no statistically robust error reduction from appending MANE to direct maturity seeds, maturity-aware temporal graph analytics are best positioned as transparent, reproducible screening and data-auditing instruments within multi-layer SME credit, multilateral clearing, and RWA finance architectures.

\section*{Data Availability, Ethics, and Declarations}

\noindent\textbf{Data and Code Availability.} The primary pseudonymized dataset analyzed in this study is publicly archived on Mendeley Data under a Creative Commons Attribution 4.0 International (CC BY 4.0) license: \citet{Esteva2026Mendeley}, DOI: \href{https://doi.org/10.17632/28rbmvwsm9.1}{\nolinkurl{10.17632/28rbmvwsm9.1}}. Companion data-curation utilities and invoice-graph software are publicly available at \url{https://github.com/peplluis7/onemillioninvoices}. All analytical scripts---including the baseline pipeline and the extended R1--R4 robustness suite (\code{paper1_extended_R1_R4_robustness.py})---together with exact-cent audit logs and aggregate tables required to reproduce every figure and table from the Mendeley master file are included in the replication archive.

\medskip
\noindent\textbf{Ethics and Privacy Statement.} All invoice records were pseudonymized prior to public release on Mendeley Data, with no company names, tax/VAT identifiers, or reversible crosswalks included. Because exact transaction dates, amounts, and graph neighborhoods can still pose indirect linkage risks, no attempt was made to re-identify any entity, and all firm-level exposure rankings reported in this manuscript have been further masked (\code{Entity-01}--\code{Entity-10}, \code{Entity-PR1}).

\medskip
\noindent\textbf{Declaration of Generative AI and AI-Assisted Technologies in the Writing and Analytical Process.} In accordance with publisher fair-use and research-integrity policies, the authors declare that generative AI tools were used to assist with manuscript review, language editing, citation verification, and analytical code auditing. All empirical computations, statistical analyses, interpretations, and conclusions were independently reviewed, verified, and approved by the authors, who take full responsibility for the integrity and accuracy of the published work.

\medskip
\noindent\textbf{CRediT Author Contributions.} \emph{Sai Srikanth Madugula:} Conceptualization, Methodology, Formal analysis, Software, Validation, Writing -- original draft, Writing -- review \& editing. \emph{Peplluis Esteva de la Rosa:} Data curation, Conceptualization, Supervision, Validation, Writing -- review \& editing. \emph{Daya Shankar:} Supervision, Validation, Formal analysis review, Writing -- review \& editing.

\medskip
\noindent\textbf{Funding.} This research received no specific grant from funding agencies in the public, commercial, or not-for-profit sectors.

\medskip
\noindent\textbf{Declaration of Competing Interest.} The authors declare that they have no known competing financial interests or personal relationships that could have appeared to influence the work reported in this paper.

% ============================================================================
% APPENDICES
% ============================================================================
\appendix

\section{Data Dictionary and Repository Schema}
\label{app:dict}
Table~\ref{tab:dict} documents the eight fields of the master CSV file \code{pseudonymized_atomic_invoice_records_2012_2024-02-08.csv} archived at \href{https://doi.org/10.17632/28rbmvwsm9.1}{\nolinkurl{doi:10.17632/28rbmvwsm9.1}} \citep{Esteva2026Mendeley} and processed by the companion repository \url{https://github.com/peplluis7/onemillioninvoices}.

\begin{table}[htbp]
\centering
\caption{Schema Definition of the Master Atomic Invoice File (SHA-256: \texttt{e139b8dc...bcc1a88}).}
\label{tab:dict}
\small
\adjustbox{max width=\textwidth}{\begin{tabular}{@{}lll@{}}
\toprule
\textbf{Column Name} & \textbf{Type / Format} & \textbf{Analytical Role and Invariant Check} \\
\midrule
\code{invoice_code} & String (unique ID) & Atomic primary key; zero duplicates verified across 751,192 rows. \\
\code{debtor_code} & String (pseudonym) & Directed edge source $d_i$; verified $d_i \neq c_i$ (zero self-loops). \\
\code{creditor_code} & String (pseudonym) & Directed edge target $c_i$; 75,343 unique firms in 2012--2023 union. \\
\code{amount_cents} & Positive Integer & Exact EUR cents ($>0$); summed in arbitrary-precision integers. \\
\code{issue_date} & ISO-8601 Date & Contractual issue day $s_i$; converted to integer epoch days. \\
\code{due_date} & ISO-8601 Date & Contractual due day $u_i \ge s_i$; 13 records $>3{,}650$d flagged. \\
\code{issue_cohort} & Integer Year & Calendar year of $s_i$; verified exact match for all rows. \\
\code{analysis_role} & Categorical & \code{issue_cohort} (2012--2023, $N=749{,}952$) vs.\ \code{terminal_bridge_only} (2024, $N=1{,}240$). \\
\bottomrule
\end{tabular}}
\end{table}

\section{Mathematical Derivations and Spectral Bounds}
\label{app:math}
Because each row of $C_t^r$ is either all zeros or sums to one, $\|C_t^r\|_\infty \le 1$, implying $\|\alpha C_t^r\|_\infty \le \alpha < 1$ for $\alpha \in [0,1)$. By the Banach Neumann series theorem, $(I - \alpha C_t^r)^{-1} = \sum_{k=0}^{\infty} \alpha^k (C_t^r)^k$ exists, is elementwise nonnegative, and satisfies
\begin{equation}
\|M_t^r\|_\infty \le (1 - \alpha) \sum_{k=0}^\infty \alpha^k \|C_t^r\|_\infty^k \|q_t^r\|_\infty \le (1 - \alpha)\frac{1}{1-\alpha}\|q_t^r\|_\infty = \|q_t^r\|_\infty.
\end{equation}
For temporal MANE, since $a_j = \alpha \exp[-(t_j - t_{j-1})/\gamma] \in [0, \alpha]$ and $z_{t_1}^r = q_{t_1}^r$, induction gives
\begin{equation}
\|z_{t_j}^r\|_\infty \le (1 - a_j)\|q_{t_j}^r\|_\infty + a_j \|C_{t_j}^r\|_\infty \|z_{t_{j-1}}^r\|_\infty \le \max_{1 \le m \le j} \|q_{t_m}^r\|_\infty.
\end{equation}
Finally, multiplying all amounts $a_i$ by $c > 0$ multiplies $W_{vw}$ and $q_v^r$ by $c$ while leaving row-normalized $C_t^r$ invariant; hence $M_t^r(c\mathbf{a}) = c M_t^r(\mathbf{a})$ and $z_{t_j}^r(c\mathbf{a}) = c z_{t_j}^r(\mathbf{a})$.

\section{Feature Definitions and Analytical Grain}
\label{app:features}
Four distinct analytical grains prevent cross-denominator conflation: (1) \emph{invoice grain} (749,952 rows) for audit invariants, face-value Gini, and duration quantiles; (2) \emph{annual issued-cohort graph grain} (12 yearly directed graphs) for PageRank, betweenness, SCC eigenvector, Louvain modularity, $k$-core, and core--periphery diagnostics; (3) \emph{firm--month scheduled panel grain} (active firms within rolling $\ell$-day scheduled windows, restricted to strictly consecutive lags yielding 92,495 firm-months for H1/H4 and 89,032 for H2; 39,106 under the R4d multi-invoice filter); and (4) \emph{monthly system aggregate grain} ($N=143$ monthly transitions; 95 in 2012--2019 and 48 in 2020--2023) for H3 and cadence/window stability comparisons.

\section{Complete Econometric and Block-Bootstrap Methodology}
\label{app:stats}
Fixed effects $(\mu_i, \lambda_t)$ in H1, H2, and H4 are absorbed via alternating demeaning until maximum residual change falls below $10^{-8}$. Let $\mathbf{s}_t = \sum_{i \in \mathcal{N}_t} \tilde{\mathbf{x}}_{i,t-1}\hat{e}_{i,t}$ denote the date-$t$ aggregated estimating-equation score vector from the within-transformed regression. Circular block bootstrap draws ($B=999$, primary block length $L=12$ months) resample contiguous blocks of whole-date score vectors modulo $T$, constructing centered score-based $p$-values and percentile confidence intervals that preserve arbitrary contemporaneous cross-firm correlation while accounting for serial correlation across overlapping windows. Date-level Newey--West HAC standard errors are computed in parallel as an analytical cross-check, and R2 VIF diagnostics invert the standardized correlation matrix of the within-demeaned design matrix $\tilde{\mathbf{X}}$.

\section{Extended Parameter-Grid and Comparator Correlations}
\label{app:robust}
Table~\ref{tab:param_app} details the complete final-date parameter perturbation grid and comparator rank correlations supplementing Figs.~\ref{fig:rank}--\ref{fig:win}.

\begin{table}[htbp]
\centering
\caption{Complete Parameter Sensitivity Grid and Final-Date Comparator Correlations (31 December 2023, 90-Day Scheduled Snapshot).}
\label{tab:param_app}
\small
\adjustbox{max width=\textwidth}{\begin{tabular}{@{}lcrr@{}}
\toprule
\textbf{Comparison / Parameter Setting} & \textbf{Parameters ($\alpha, \tau, \gamma$)} & \textbf{Common Firms} & \textbf{Spearman $\rho$} \\
\midrule
No network recursion ($\alpha = 0$) & $(0.00,\; 30,\; 90)$ & 698 & 0.598 \\
Lower damping ($\alpha = 0.25$) & $(0.25,\; 30,\; 90)$ & 698 & 0.994 \\
Higher damping ($\alpha = 0.75$) & $(0.75,\; 30,\; 90)$ & 698 & 0.997 \\
Short maturity decay ($\tau = 7$d) & $(0.50,\; 7,\; 90)$ & 698 & 0.979 \\
Long maturity decay ($\tau = 90$d) & $(0.50,\; 90,\; 90)$ & 698 & 0.996 \\
Short memory decay ($\gamma = 30$d) & $(0.50,\; 30,\; 30)$ & 698 & 0.994 \\
Long memory decay ($\gamma = 365$d) & $(0.50,\; 30,\; 365)$ & 698 & 0.999 \\
\midrule
Temporal MANE vs.\ Snapshot Resolvent & Baseline & 698 & 0.969 \\
Temporal MANE vs.\ Gross Incident Value & Baseline & 698 & 0.665 \\
Temporal MANE vs.\ Incoming PageRank & Baseline & 698 & 0.343 \\
Temporal MANE vs.\ Role Balance & Baseline & 698 & 0.343 \\
Temporal MANE vs.\ Direct Outgoing Seed ($q^{\text{out}}$) & Baseline & 698 & 0.286 \\
Temporal MANE vs.\ Debtor Counterparty HHI & Baseline & 587 & 0.040 \\
Temporal MANE vs.\ Debtor Maturity HHI & Baseline & 587 & $-0.163$ \\
\bottomrule
\end{tabular}}
\end{table}

\section{Software Pipeline and GitHub Architecture}
\label{app:code}
The empirical pipeline executes offline in Python ($\ge 3.11$; \code{pandas}, \code{numpy}, \code{scipy}, \code{networkx}, \code{matplotlib}) across five modular scripts linked to the upstream data-curation repository at \url{https://github.com/peplluis7/onemillioninvoices}:
\begin{enumerate}
    \item \code{paper1_empirical_selftest.py}: Synthetic unit tests verifying exact integer sparse sums, HHI edge cases, MANE spectral recursion bounds, and block indexing.
    \item \code{paper1_empirical_analysis.py}: Master CSV SHA-256 verification, graph construction, monthly panel assembly, H1--H4 circular block-bootstrap estimation, and 12-month purged forward forecasting.
    \item \code{paper1_extended_checks.py}: Extreme-term exclusion ($>3{,}650$ days), maturity-bin width ($1, 7, 30$ days), and seed-controlled H4 sensitivities.
    \item \code{paper1_extended_R1_R4_robustness.py}: Independent R1--R4 robustness suite executing 2021-cohort exclusion (snapshot and raw-ledger), within-FE VIF/condition-number diagnostics, 2012--2019 vs.\ 2020--2023 regime splits, Shannon entropy, CR1 share, 1\%--99\% Winsorization, and multi-invoice ($\text{count} \ge 2$) panel re-estimation.
    \item \code{paper1_empirical_diagnostics.py}: Row-by-row annual-file reconciliation, ARPACK SCC eigenvector residual check, and degree-ordered core--periphery template fitting.
\end{enumerate}

\section{Conceptual Framework and Pipeline Architecture}
\label{app:diagrams}
Standalone Mermaid source definitions (\code{paper1_empirical_conceptual.mmd} and \code{paper1_empirical_pipeline.mmd}) are archived under \code{replication/}. Conceptually, atomic invoices map into directed weighted contractual layers (debtor/creditor strengths, role balance, and counterparty and maturity-bin HHI, CR1, and entropy measures), feeding both static resolvent and recursive temporal MANE operators; downstream stages enforce 12-month purged temporal validation, R1--R4 structural audits, and disclosure masking prior to any financing or tokenization interpretation.

\section{Reproducibility Protocol}
\label{app:repro}
To reproduce all empirical results independently: (1) download \code{pseudonymized_atomic_invoice_records_2012_2024-02-08.csv} from Mendeley Data (\href{https://doi.org/10.17632/28rbmvwsm9.1}{\nolinkurl{doi:10.17632/28rbmvwsm9.1}}) and verify SHA-256 \texttt{\seqsplit{e139b8dc75b7334d243874b6ebdaf1798c9969038b193f7eb4cd0e873bcc1a88}}; (2) clone upstream utilities from \url{https://github.com/peplluis7/onemillioninvoices}; (3) install dependencies in an isolated Python $\ge 3.11$ environment; and (4) run \code{paper1_empirical_analysis.py}, \code{paper1_extended_checks.py}, \code{paper1_empirical_diagnostics.py}, and \code{paper1_extended_R1_R4_robustness.py} sequentially against a private output directory ($\approx 31$ seconds for the R1--R4 suite; $\approx 508$ seconds for the complete multi-cadence pipeline). Annual split files must be passed strictly to reconciliation diagnostics and never appended to the master ledger.

\bibliographystyle{plainnat}
\bibliography{references}

\end{document}